\documentclass[preprint,sort&compress,12pt]{elsarticle}

\usepackage{graphicx}
\usepackage{amssymb}
\usepackage{amsthm}
\usepackage{bbm}
\usepackage{lineno}
\usepackage{fullpage}
\usepackage[colorinlistoftodos]{todonotes}
\usepackage{url}
\usepackage{listings}
\usepackage[colorlinks=true]{hyperref}

\usepackage{lineno}
\usepackage{subfig}
\usepackage{color,soul}
\usepackage{enumitem}
\usepackage{mathtools}

\definecolor{lightblue}{rgb}{.90,.95,1}
\definecolor{darkgreen}{rgb}{0,.5,0.5}

\journal{Elsevier}

\begin{document}

\begin{frontmatter}



 \title{A Random Matrix Approach for Quantifying Model-Form Uncertainties in Turbulence Modeling}



 \author[vt]{Heng Xiao\corref{corxh}}
 \cortext[corxh]{Corresponding author. Tel: +1 540 231 0926}
 \ead{hengxiao@vt.edu}
 \author[vt]{Jian-Xun Wang}
\author[usc]{Roger G. Ghanem}
\address[vt]{Department of Aerospace and Ocean Engineering, Virginia Tech, Blacksburg, VA 24060, United States}
\address[usc]{Department of Aerospace and Mechanical Engineering, University of Southern California,
  Los Angeles, CA 90089,  United States}

\begin{abstract}
  With the ever-increasing use of Reynolds-Averaged Navier--Stokes (RANS) simulations in
  mission-critical applications, the quantification of model-form uncertainty in RANS models has
  attracted attention in the turbulence modeling community. Recently, a physics-based, nonparametric
  approach for quantifying model-form uncertainty in RANS simulations has been proposed, where
  Reynolds stresses are projected to physically meaningful dimensions and perturbations are
  introduced only in the physically realizable limits (Xiao et al., 2015. Quantifying and reducing
  model-form uncertainties in Reynolds-averaged Navier--Stokes simulations: An open-box,
  physics-based, Bayesian approach, arXiv:1508.06315).  However, a challenge associated with this
  approach is to assess the amount of information introduced in the prior distribution and to avoid
  imposing unwarranted constraints.  In this work we propose a random matrix approach
    for quantifying model-form uncertainties in RANS simulations with the realizability of the
    Reynolds stress guaranteed, which is achieved by construction from the Cholesky factorization of the
    normalized Reynolds stress tensor.  Furthermore, the maximum entropy principle is used to
  identify the probability distribution that satisfies the constraints from available information
  but without introducing artificial constraints.  We demonstrate that the proposed approach is able
  to ensure the realizability of the Reynolds stress, albeit in a different manner
  from the physics-based approach. Monte Carlo sampling of the obtained probability distribution is
  achieved by using polynomial chaos expansion to map independent Gaussian random fields to the
  Reynolds stress random field with the marginal distributions and correlation structures as
  specified.  Numerical simulations on a typical flow with separation have shown physically
  reasonable results, which verifies the proposed approach. Therefore, the proposed method is a
  promising alternative to the physics-based approach for model-form uncertainty quantification of
  RANS simulations. The method explored in this work is general and can be extended to other complex
  physical systems in applied mechanics and engineering.
\end{abstract}

\end{frontmatter}

\section*{Notations}
We summarize the convention of notations below because of the large number of symbols used in this
paper. The general conventions are as follows:
\begin{enumerate} 
\item Upper case letters with brackets (e.g., $[R]$) indicate matrices or tensors; lower case
  letters with arrows (e.g., $\vec{v}$) indicate vectors; undecorated letters in either upper or
  lower cases indicate scalars.  An exception is the spatial coordinate, which is denoted as $x$ for
  simplicity but is in fact a $3 \times 1$ vector.  Tensors (matrices) and vectors are also
  indicated with index notations, e.g., $R_{ij}$ and $v_i$ with $i, j = 1, 2, 3$. In this paper, $i$
  and $j$ are used with tensor indices while $\alpha$ and $\beta$ are used as general indexes, e.g.,
  for the modes of Kahunen--Loeve expansion or for the terms in the polynomial chaos expansion.
 
\item Bold letters (e.g., $[\mathbf{R}]$) indicate random variables (including scalars, vectors, and
  matrices), the non-bold letters (e.g., $[R]$) indicate the corresponding realizations, and
  underlined letters (e.g., $[\underline{R}]$) indicate the mean.

\item Symbols $\mathbb{M}_d^{s}$, ${M}_d^{+}$, and $\mathbb{M}_d^{+0}$ indicate the sets of
  symmetric, symmetric positive definite, and symmetric positive semi-definite matrices,
  respectively, of dimension $d \times d$ with the following relation: $\mathbb{M}_d^{s} \subset
  {M}_d^{+} \subset \mathbb{M}_d^{+0}$.
\end{enumerate}
This work deals with Reynolds stresses, which are rank two tensors. Therefore, it is implied
throughout the paper that all random or deterministic matrices have sizes $3 \times 3$ with real
entries unless noted otherwise.  Finally, a list of nomenclature is presented in~\ref{app:notation}.

\section{Introduction}
\label{sec:intro}

Numerical models based on Reynolds-Averaged Navier--Stokes (RANS) equations are the dominant tool
for the prediction of turbulent flows in industrial and natural processes.  For many flows the
predictions from RANS simulations have large uncertainties, which are mostly attributed to the
phenomenological closure models for the Reynolds
stresses~\cite{pope2000turbulent,oliver2009uncertainty}. Previous efforts in quantifying and reducing
model uncertainties in RANS simulation have mostly followed parametric approaches, e.g., by
perturbing, tuning, or inferring the model parameters in these closure
models~\cite{margheri2014epistemic}. Recently, researchers in the turbulence modeling community have
recognized the limitations of the parametric approach and started exploring nonparametric approaches
where uncertainties are directly injected into the Reynolds
stresses~\cite{oliver2009uncertainty,emory2011modeling,emory2013modeling,gorle2013framework,xiao-mfu}.

In their pioneering work, Iaccarino and co-workers~\cite{emory2011modeling,emory2013modeling}
projected the Reynolds stress onto the six physically meaningful dimensions (its magnitude, shape,
and orientation) and perturbed the Reynolds stresses towards the limiting states in the physically
realizable range.  Xiao et al.~\cite{xiao-mfu} utilized the six physical dimensions as the basis to
parameterize the Reynolds stress uncertainty space, and the physical-based parameterization is
further used for systematic exploration of the space and for Bayesian inferences to incorporate
observation data. However, although the physics-based parameterization allows for a full
representation of the uncertainty space, two limitations need to be addressed. First, the
realizability constraints on Reynolds stresses provide guidance only on the shape of the Reynolds
stress (in the form of the classical Lumley triangle~\cite{lumley1977return} or the recently
proposed Barycentric triangle~\cite{banerjee2007presentation}).  In comparison, much weaker
constraints can be imposed on the magnitude and the orientation, except for the constraint that the
magnitude (i.e., turbulent kinetic energy) must be positive.  For this reason and partly due to
stability considerations, previous works on injecting uncertainty to the physical projections of the
Reynolds stress have focused on the shape and
magnitude~\cite{emory2011modeling,xiao-mfu,gorle2013framework}, and they did not introduce
uncertainties to the orientations.  Second, in the context of Bayesian inference of Reynolds stress
uncertainties as pursued by Xiao and co-workers~\cite{xiao-mfu, mfu2, mfu3}, it is not
straightforward to specify a prior distribution on the physical variables. Xiao et
al.~\cite{xiao-mfu} specified Gaussian priors for the Reynolds stress discrepancy in terms of the
tensor shape parameters and the logarithmic of the magnitude. In principle the same prior can be
specified for the orientation angles.  However, it is not clear how much information is introduced
into the prior with this choice of probability distributions (i.e., log-normal distribution for the
magnitude and standard normal for others). Moreover, in lack of physical insight, it is not clear how
large the variance for each physical variable should be compared to each other.  Although the
prior plays a minor role in the Ensemble Kalman filtering based inference with a moderate
amount of data as studied in Xiao et al.~\cite{xiao-mfu}, it can be of critical importance for pure
uncertainty propagation as pursued by Iaccarino and co-workers
\cite{emory2011modeling,emory2013modeling,gorle2013framework} and for Bayesian inferences with small
amounts of data. In particular, it is critical that no artificial constraints are introduced with
overly confident prior distributions.

The entropy measure of information has been proposed more than half a century
ago~\cite{shannon1948math}. Since then, the maximum entropy principle has been extensively used as
the guideline to specify prior probability distributions for Bayesian
inferences~\cite{jaynes57info}. The principle states that out of all possible distributions that
satisfy the constraints from available information, the probability distribution that has the
maximum entropy is a good prior distribution.  Since entropy is a measure of randomness, the maximum
entropy distribution is the most non-committal and most random in the dimensions in which no
information is available.  Consequently, it introduces the least amount of information in addition
to the specified constraints. While the maximum entropy principle has been used extensively in many
disciplines such as communications~\cite{coifman1992entropy} and image
processing~\cite{zhu1998filters}, the use of maximum entropy principle in conjunction with random
matrix theory for quantifying model-form uncertainties in physical systems is only a recent
development. Soize~\cite{soize2000nonparametric} was the first to derive the maximum entropy
probability distribution of a symmetric positive definite random matrices with a specified mean
field. Applications to structural vibration problems were demonstrated, where the mass, stiffness,
and damping matrices of the real system are described as random matrices with the corresponding
matrices in the reduced model as their means.  The framework was further extended to nonlinear
structural dynamics problems and to other applications (e.g., composite
materials~\cite{das2009bounded}, porous media\cite{guilleminot2012stochastic}) and to problems with
more complicated constraints, (e.g., on the variance of eigenvalues~\cite{mignolet2008nonparametric,
  guilleminot2012stochastic}, as well lower and upper bounds on the
matrices~\cite{das2009bounded}). See~\cite{soize2005random,soize2013stochastic} for comprehensive
reviews of the recent development.

Since Reynolds stresses are symmetric positive definite tensors, it seems natural to use random
matrix approach to describe the Reynolds stresses for the quantification of RANS modeling
uncertainties. However, the authors' are not aware of any existing work in the literature that
applied the random matrix theory in quantifying model-form uncertainties in RANS simulations.  While
many of the theories developed by Soize and co-workers can be applied here straightforwardly, an
important characteristics of the RANS modeling application is that the Reynolds stresses are
described by a random matrix field with spatial correlations. In contrast, the matrices in the
previously investigated applications either involve only a few individual matrices (e.g., the mass,
stiffness, and damping matrices in structural dynamics applications~\cite{soize2000nonparametric} )
or a large number of random matrices without spatial correlation structure (e.g., the effective
constitutive matrices in the meso-scale modeling of composite materials~\cite{das2009bounded}).  A
notable exception is the recent work of Guilleminot~\cite{guilleminot2012stochastic}, where the
permeability tensor is modeled as a random matrix field with a specified correlation structure.

In the present contribution, we use an approach based on the random matrix theory and the maximum
entropy principle to quantify model-form uncertainties in RANS modeling. The objectives of this work
are two-fold. First, the proposed framework for quantifying model form uncertainties in RANS
simulations provides an alternative to the physics-based approach investigated by Iaccarino and
co-workers~\cite{emory2011modeling,emory2013modeling,gorle2013framework} and Xiao and
co-workers~\cite{xiao-mfu,mfu2,mfu3}. The advantages and disadvantages of both approaches can thus
be contrasted and compared later on.  Despite the better mathematical rigorousness of the maximum
entropy approach, it is possible that the physics-based approach may still be preferred by RANS
simulations practitioners for both uncertainty quantification and Bayesian inferences. This is
because of its convenience in incorporating physical prior knowledge.  As such, another objective
and motivation for this work is to provide basis for gauging the departure of the priors used in the
physics-based approach from the maximum entropy condition.  By comparing the prior distributions of
the obtained Reynolds stress from both approaches, the current framework can reveal the amount of
information introduced in the physics-based prior distribution, and thus provide guidance for the
choice of priors in the physics-based approach.

The rest of the paper is organized as follows. Section~2 introduces the realizability of Reynolds
stresses in details and argues that the current framework of modeling Reynolds stress as positive
semidefinite tensors can guarantee realizability in the same way as in the previously proposed
physics-based approaches~\cite{xiao-mfu}. Section~3 introduces the random matrix framework for
model-form uncertainty in RANS simulation and discusses its implementations. Section~4 uses the flow
over periodic hills as an example application to demonstrate the performance of the proposed
method. The probability measure of the obtained distribution on the limiting states is discussed in
Section 5.  Finally, Section~6 concludes the paper.

\section{Realizability of Reynolds Stresses: Physical vs.\ Mathematical Perspectives}

\subsection{The Origin and History of Realizability Constraints}
\label{sec:realize-origin}

For incompressible flows, the RANS equations are obtained by decomposing the instantaneous velocity
$\mathbf{v}_i$, into the mean quantity $\underline{v}_i$ and its fluctuation $\mathbf{v}_i'$. That
is, $\mathbf{v}_i = \underline{v}_i + \mathbf{v}_i'$. Note the bold letters used to denote the
instantaneous velocity, indicating that it is considered a random variable. Substituting the
decomposition above into the Navier Stokes equation leads to a covariance term $\langle
\mathbf{v}_i' \mathbf{v}_j' \rangle$ of the velocity fluctuations. This term is referred to as
Reynolds stress and is denoted as $R_{ij}$ for simplicity\footnote{This is actually the negative of
  the Reynolds stress~\cite{popebook}.}, but the covariance nature of this term has profound
implications. Specifically, a Reynolds stress tensor must be symmetric positive semi-definite, i.e.,
$R_{ij} \in \mathbb{M}_3^{+0}$, where $\mathbb{M}_3^{+0}$ is the set of symmetric positive
semidefinite matrices. In fact, the Reynolds stresses are mostly positive definite (i.e., all
eigenvalues are positive), i.e., $R_{ij} \in \mathbb{M}_3^+$, and zero eigenvalues occur only in
extreme cases, e.g., on the wall boundaries or in turbulence-free regions, where velocity
fluctuations for one or more components are zero.  For practical purposes it is not essential to
distinguish between $\mathbb{M}_3^{+}$ and $\mathbb{M}_3^{+0}$ when quantifying model form
uncertainties in RANS simulations.

For a tensor to be a valid Reynolds stress tensor, or equivalently, to be physically realizable,
there must exist a velocity that has this covariance. The physical realizability of Reynolds
stresses and the associated second-order closure models were a topic of intensive research in the
early years of turbulence model development.  Early works of
Schumann~\cite{schumann1977realizability}, Lumley~\cite{lumley1978computational}, and
Pope~\cite{pope1985pdf}, among others, have lead to a class of realizable second-order closure
models~\cite{speziale1994realizability}. Since these models solve an evolution equation for the
Reynolds stresses, these models guarantee the realizability by imposing certain constraints on the
rate of change $[\dot{R}]$ of the Reynolds stress, i.e., the constraint that its principle stress
components (e.g., eigenvalues) do not further decrease when at zero~\cite{pope1985pdf}.  The well
known Lumley triangle~\cite{lumley1977return} provided a map for all realizable turbulence by
projecting the Reynolds stress to a plane. Specifically, the following decomposition is performed on
the Reynolds stress:
\begin{equation}
  \label{eq:tau-decomp}
  [R] = 2 k \left( \frac{1}{3} [I] +  [A] \right)
  = 2 k \left( \frac{1}{3} [I] + [E] [\Lambda] [E]^T \right)
\end{equation}
where $k$ is the turbulent kinetic energy which indicates the magnitude of $[R]$; $[I]$ is the
second order identity tensor; $[A]$ is the anisotropy tensor; $[E] = [\vec{e}_1, \vec{e}_2,
\vec{e}_3]$ and $[\Lambda] = \textrm{diag}[\tilde{\lambda}_1, \tilde{\lambda}_2, \tilde{\lambda}_3]$
where $\tilde{\lambda}_1+ \tilde{\lambda}_2 + \tilde{\lambda}_3=0$ are the orthonormal eigenvectors
and the corresponding eigenvalues of $[A]$, respectively, indicating the shape and orientation of
$[R]$.  The Lumley triangle is thus defined on a plane with the second and third invariants
($\textrm{II}$ and $\textrm{III}$) of the anisotropy tensor as coordinates, where $\textrm{II} =
(\operatorname{tr}[A])^2 - \operatorname{tr}([A]^2) $ and $\textrm{III}=\det[A]$.

The realizable turbulence models have demonstrated superior performance in some cases compared to
their standard counterpart models that do not guarantee realizability. Interestingly, however,
several decades after their development they have not dominated the non-realizable models. Among the
handful of standard, most widely used turbulence models (e.g.,
$k$--$\varepsilon$~\cite{launder1974numerical}, $k$--$\omega$ SST model~\cite{menter1994two}, and SA
model~\cite{spalart94SA}) that are implemented in industrial standard CFD packages, none of them are
realizable models.

\subsection{Realizability in Physics-Based Model-Form Uncertainty Quantification}
\label{sec:physics-proj}
\begin{figure}[!htbp]
  \centering
   \subfloat[]{\includegraphics[width=0.5\textwidth]{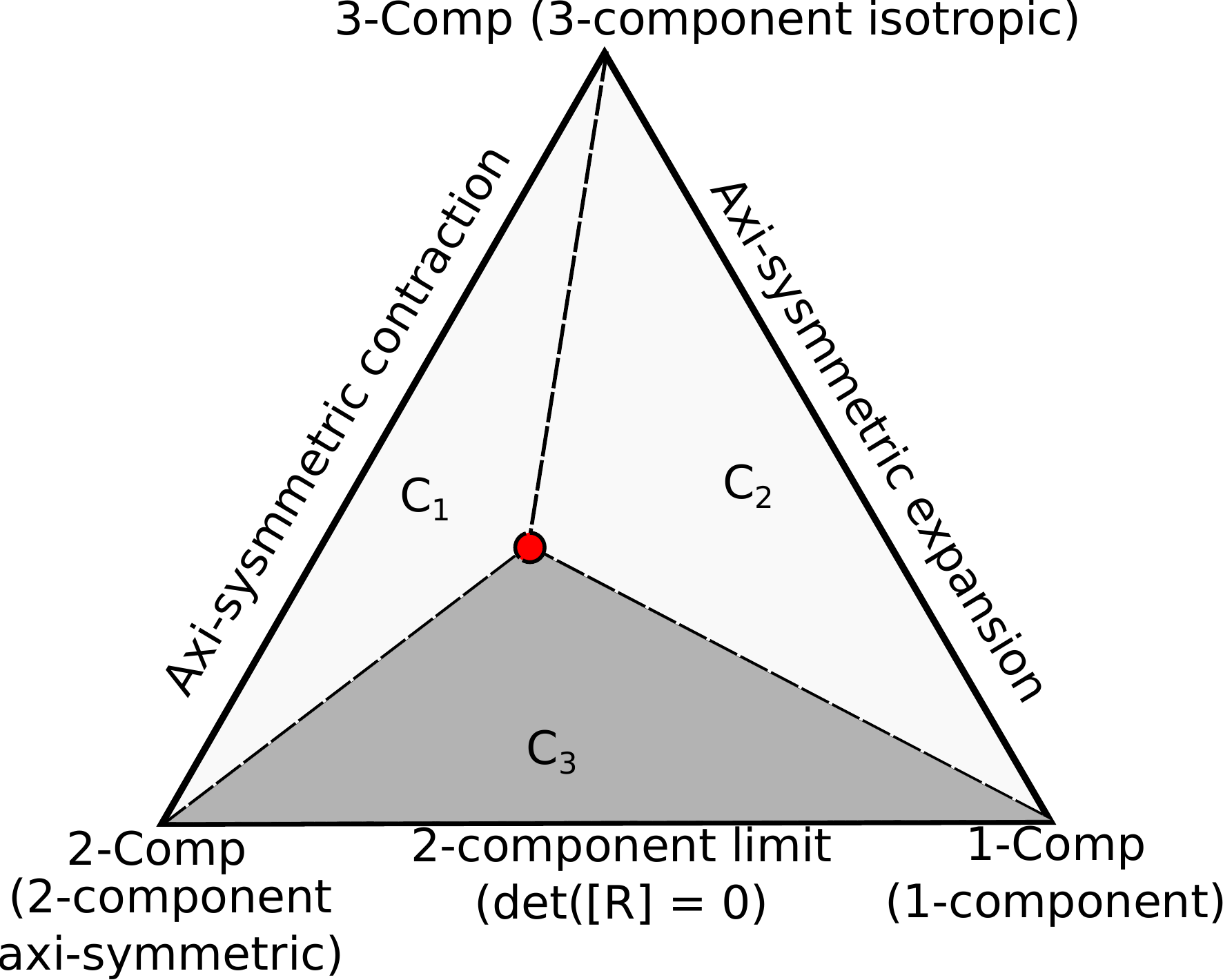}}
   \subfloat[] {\includegraphics[width=0.5\textwidth]{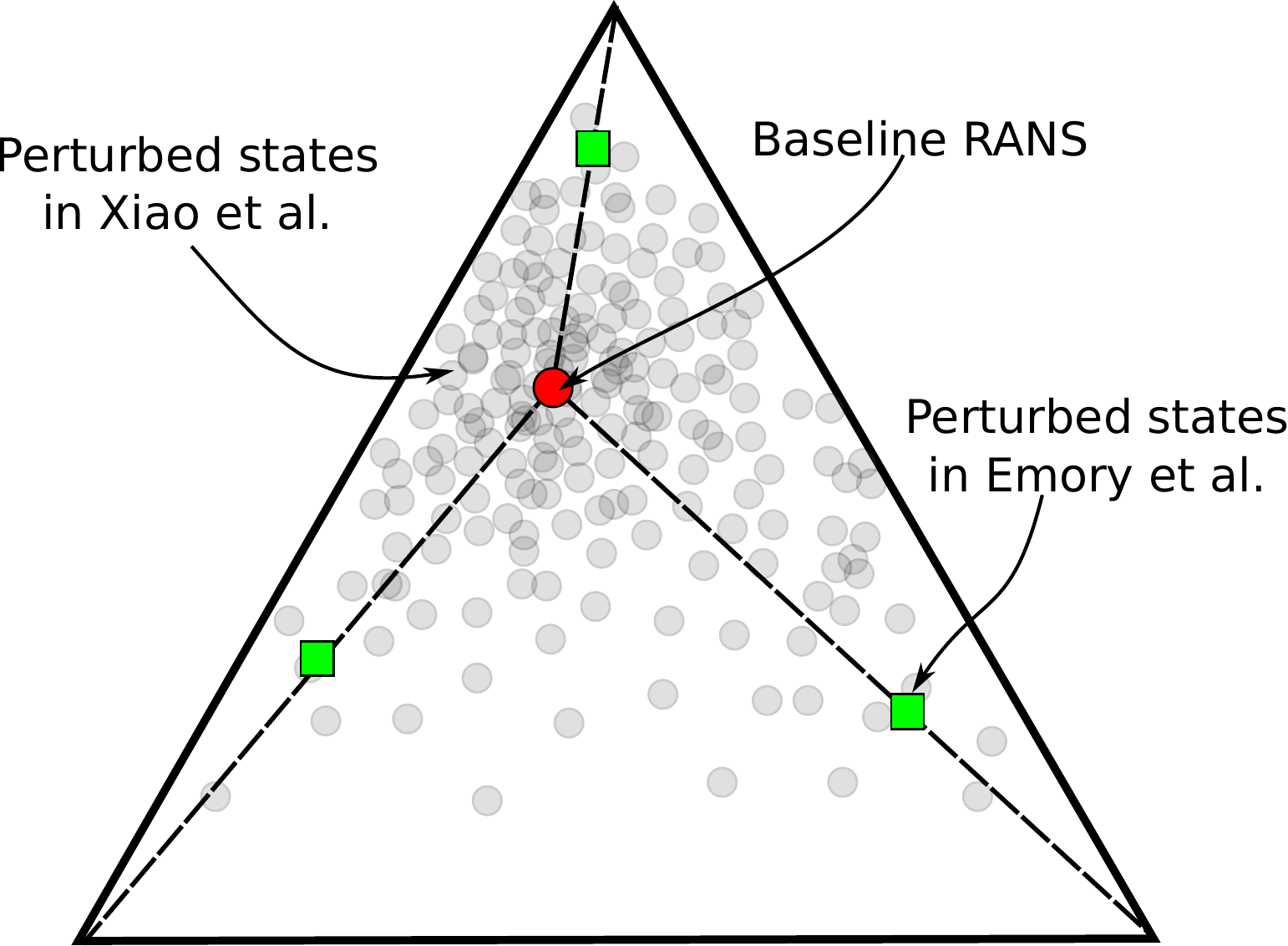}}
   \caption{ (a) Barycentric triangle as a way of delineating realizable turbulence, its definition
     and physical interpretation. Elements in the set $\mathbb{M}_d^{+0}$ of positive semidefinite
     matrices maps to the interior and edges of the triangle, while the set
     $\mathbb{M}_d^{+0}-\mathbb{M}_d^{+}$ of singular matrices maps (whose element have zero
     determinants) maps to the bottom edge.  (b) Model-form uncertainty quantification through
     perturbation of Reynolds stresses within the physically realizabile limit enclosed by the
     Barycentric triangle.  The different schemes of Emory et al.~\cite{emory2011modeling} and Xiao
     et al.~\cite{xiao-mfu} are shown.}
  \label{fig:bary}
\end{figure}

The past two decades have not seen significant literature on the realizability of turbulence models,
partly because the research on RANS model development has been stagnant all together.  However,
recently the pioneering research of Iaccarino and
co-workers~\cite{emory2011modeling,emory2013modeling, gorle2014deviation} on the quantification of
model form uncertainties in RANS simulations has revived the interests on Reynolds stresses
realizability.  Specifically, they performed the decomposition as shown in Eq.~(\ref{eq:tau-decomp})
and mapped the eigenvalues $\tilde{\lambda}_1$, $\tilde{\lambda}_2$, and $\tilde{\lambda}_3$ to
Barycentric coordinates $(C_1, C_2, C_3)$ as follows:
\begin{subequations}
  \label{eq:lambda2c}
\begin{align}
  C_1 & = \tilde{\lambda}_1 - \tilde{\lambda}_2 \\
  C_2 & = 2(\tilde{\lambda}_2 - \tilde{\lambda}_3) \\
  C_3 & = 3 \tilde{\lambda}_3 + 1 \ .
\end{align}  
\end{subequations}
As illustrated in Fig.~\ref{fig:bary}a, the Barycentric coordinates ($C_1$, $C_2$, $C_3$) of a point
indicate portion of areas of the three sub-triangles formed by the point and the edge labeled as
$C_1$, $C_2$, and $C_3$, respectively, in the Barycentric triangle.  For example, the ratio between
the shaded sub-triangle and the entire triangle is $C_3$. A point located on the top vertex
corresponds to $C_3 = 1$ while a point located on the bottom edge (labeled $C_3$ in
Fig.~\ref{fig:bary}a) has $C_3$ = 0.  Therefore, each coordinate ranges from 0 to 1, and they sum to
1 for any point, i.e., $C_1 + C_2 + C_3 = 1$.  The three edges labeled $C_1$, $C_2$ and $C_3$ are
opposite to the vertices representing one-component, two-component, and three-component limits of
realizable turbulence states. The Barycentric coordinates have clear physical interpretation in that
they indicate the componentality (i.e., dimensionality) of the turbulence.  For example, the upper
corner in Fig.~\ref{fig:bary}a, which corresponds to $C_3 = 1$ or equivalently a Barycentric
coordinate of $(0, 0, 1)$, indicates isotropic, fully three-dimensional turbulence.  Detailed
physical interpretation of the vertices and edges are shown in Fig.~\ref{fig:bary}a. The Barycentric
triangle is similar to the Lumley triangle in that it also encloses all realizable turbulence
states.  Emory et al.~\cite{emory2011modeling} estimated the uncertainties in RANS simulations by
perturbing the Reynolds stress towards the three limiting states, i.e., the vertices of the
Barycentric triangle. This is illustrated in Fig.~\ref{fig:bary}b as squares. Based on their work,
Xiao et al.~\cite{xiao-mfu} further mapped the Barycentric coordinates to natural coordinates, on
which the equilateral triangle is mapped to a unit square. They parameterized the uncertainty space
on the natural coordinates and systematically explored the uncertainty space. The samples as
obtained in Xiao et al.~\cite{xiao-mfu} are illustrated in Fig.~\ref{fig:bary}b, which are in
contrast to the three perturbed states of Emory et al.~\cite{emory2011modeling}.

While both Emory et al.~\cite{emory2011modeling} and Xiao et al.~\cite{xiao-mfu} injected
uncertainties only to the magnitude and shape of the Reynolds stress tensor, it is theoretically
possible to perturb the orientation as well.  This work is not directly concerned with the
physics-based approach for uncertainty quantification, but the physics-based parameterization will
facilitate interpretations of the samples obtained with the random matrix approach.  To this end, a
parameterization scheme for the orthonormal eigenvectors $E =[\vec{e}_1, \vec{e}_2, \vec{e}_3]$ is
needed.  We use the Euler angle with the $z$-$x'$-$z''$ convention to describe the orientation of
the Reynolds stress tensor~\cite{goldstein80euler}. That is, if a local coordinate system
$x$--$y$--$z$ spanned by the three eigenvectors of $[R]$ was initially aligned with the global
coordinate system ($X$--$Y$--$Z$), the current configuration could be obtained by the following
three consecutive intrinsic rotations about the axes of the local coordinate system: (1) a rotation
about the $z$ axis by angle $\varphi_1$, (2) a rotation about the $x$ axis by $\varphi_2$, and (3)
another rotation about its $z$ axis by $\varphi_3$. Note that the local coordinate axes may change
orientations after each rotation.

In summary, the Reynolds stress can be represented with six independent variables with clear
physical interpretations. The turbulent kinetic energy $k$ represents its magnitude; the Barycentric
coordinates $C_1, C_2, C_3$, of which only two are independent, indicate its shape; the Euler angles
$\varphi_1, \varphi_2$ and $\varphi_3$ represent its orientation.  Realizability can be guaranteed
by injecting uncertainties in these physical variables, a fact that has been exploited in the
physics-based approach for RANS model-form uncertainty quantification investigated in earlier works.

\subsection{Mathematical and Physical Interpretations of Realizability}
\label{sec:equiv}
The physics-based approach based on Barycentric triangle as explored by Emory et
al.\cite{emory2011modeling} and Xiao et al.~\cite{xiao-mfu} certainly has clear advantages.
However, it is not the only way to guarantee realizability. Here we claim that an alternative method
to guarantee the Reynolds stress realizability in the context of RANS model-form uncertainty
quantification is to model $[R]$ as a random matrix defined on the set $\mathbb{M}_3^{+0}$ of
positive semidefinite matrices. In other words, we can directly draw $[R]$ from the set
$\mathbb{M}_3^{+0}$ and all obtained samples are guaranteed to be realizable Reynolds stresses.  To
support this claim, we utilize a theorem on the equivalence between the set of positive semidefinite
matrices and the set of co-variance matrices (see e.g., \cite{stefanica14financial}).  That is, a
covariance matrix $[R]$ must be positive semidefinite; conversely, every symmetric positive
semi-definite matrix $[R]$ is a covariance matrix. The proof for the first part is straightforward
and is thus omitted. However, the proof for the second part is of central importance to the theme of
our work, and thus it is reproduced below.  We construct a vector $\vec{\mathbf{v}} = [R]^{1/2} \,
\vec{\mathbf{b}}$ with $[R]^{1/2}$ being the symmetric square root of $[R]$ and $\vec{\mathbf{b}}$
has a covariance of identity matrix, i.e., $\operatorname{Cov} \{\vec{\mathbf{b}},
\vec{\mathbf{b}}\} = [I]$, where $\operatorname{Cov} \{\cdot\}$ indicates covariance. It can be seen
that $\operatorname{Cov}\{R^{1/2} \vec{\mathbf{b}}, R^{1/2} \vec{\mathbf{b}}\} = [R]^{1/2} \,
\operatorname{Cov}\{\vec{\mathbf{b}}, \vec{\mathbf{b}}\} \, [R]^{1/2} = [R]$.  For any sample $[R]$
drawn from the set $\mathbb{M}_d^{+0}$ of positive semidefinite matrices, there is a real-valued
velocity vector $\vec{\mathbf{v}}$ whose covariance is $[R]$. Recalling the definition and origin of
the Reynolds stress realizability described in Section~\ref{sec:realize-origin}, it can be seen that
the matrix $[R]$ is a realizable Reynolds stress. Therefore, it is concluded that the
  mathematical constraint that $[R]$ must belong to the set $\mathbb{M}_3^{+0}$ can equally
  guarantee the realizability of the obtained Reynolds stresses, albeit in a different manner from
  the physics-based approach. The latter utilized the constraint that the projection of Reynolds
  stresses must reside in the Barycentric triangle to ensure the Reynolds stress realizability.

We conclude this section with the following three \emph{equivalent} statements on the realizability
constraints of the Reynolds stresses:
\begin{enumerate}
\item a Reynolds stress must reside within or on the edge of the Lumley triangle or Barycentric
  triangle after certain transformations~\cite{banerjee2007presentation};
\item a Reynolds stress tensor must be positive semidefinite; and
\item a Reynolds stress tensor must be the covariance matrix of a real-valued vector, i.e., a
  velocity vector.
\end{enumerate}
The equivalence between realizability and the invariants mapping within Lumley triangle was
established by Lumley and Neuman~\cite{lumley1977return} and stated in ref.~\cite{popebook}
(pp. 394); the equivalence between the positive semidefinite requirement on the Reynolds stress
and the realizability conditions was proved by Schumann~\cite{schumann1977realizability}.  The
relation among three statements above is a pivotal element for anchoring the present contribution to
the existing literature of turbulence
modeling~\cite{schumann1977realizability,lumley1977return,emory2011modeling}.  However, note that
the proposed random matrix approach for quantifying RANS model-form uncertainty is fundamentally
different from the physics-based approach developed in ref.~\cite{xiao-mfu}. The most important
difference lies in the fact that in the physics-based approach the user specifies the probability
distribution on the physical variables, while in the random matrix approach the maximum entropy
principle provides the basis for computing the distribution of the matrices.



\section{Random Matrix Approach and Its Implementation}

\subsection{Probability Model for Reynolds Stress Tensors}
\label{sec:pdf-1pt}
Based on the discussions above on the nature of Reynolds stresses, we build a
probabilistic model for the Reynolds stress tensors. The development below mostly follow the works of
Soize, Ghanem, and co-workers~\cite[e.g.,][]{soize2000nonparametric,das2009bounded}.

Using $\mathbb{M}_d^{s}$ to denote the set of all symmetric matrices, the measure of matrix $[R] \in
\mathbb{M}_d^{s}$ is given as~\cite{soize2005random}:
 \begin{equation}
   \label{eq:muda}
   d[R] = 2 ^{d(d-1)/4} \prod_{1 \le i \le j  \le d} dR_{ij}
 \end{equation}
 where $\prod$ indicates product, $R_{ij}$ are individual element of matrix $[R]$, and $dR_{ij}$ is
 the Lebesgue measure on the set $\mathbb{R}$ of real numbers. Since Reynolds stresses are rank two
 tensors, $d = 3$ is implied.

 The objective is to define a probability measure $p_{[\text{\textbf{\tiny R}}]} : \mathbb{M}_d^{+0}
 \mapsto \mathbb{R}^+$ on the symmetric matrices set that maps a symmetric, positive semidefinite,
 second-order random matrix $[\mathbf{R}]$ to a positive number. While such a measure is not unique,
 in this work we seek a probability density function (PDF) that has the maximum
 entropy. Specifically, let $P_{ad}$ be the set of all the PDFs from $\mathbb{M}_d^{+0}$ to
 $\mathbb{R}^+$, the maximum entropy principle states that the most non-committal PDF is the one
 such that all available constraints are satisfied but without introducing any other artificial
 constraints~\cite{guilleminot2012stochastic}, i.e.,
\begin{equation}
  \label{eq:maxent}
  p_{[\text{\textbf{\tiny R}}]} =  \operatorname{arg} \max_{p \in P_{ad}} S(p)
\end{equation}
where the measure of entropy $S(p)$ of the PDF $p$ is defined as~\cite{soize2000nonparametric}:
\begin{equation}
  \label{eq:entropy}
  S(p) = - \int_{\mathbb{M}_d^{+0}} p([R]) \ln p([R]) d[R]  
\end{equation}
where $\ln$ denotes the natural logarithm.

In the context of RANS-based turbulence modeling, the following constraints should be satisfied by
the PDF of Reynolds stress tensors:
\begin{enumerate}
\item All realizations $[R]$ of the Reynolds stress $[\mathbf{R}]$ must be physically
  realizable. This constraint is automatically satisfied by defining $[\mathbf{R}]$ on the set
  $\mathbb{M}_d^{+0}$, i.e., $[\mathbf{R}] \in \mathbb{M}_d^{+0}$. 
\item Normalization condition for the PDF:
  \begin{equation}
    \label{eq:normalization}
    \int_{\mathbb{M}_d^{+0}} p_{[\text{\textbf{\tiny R}}]}([R]) d[R]   = 1 ,
  \end{equation}
  which is self-evident.
\item We assume that a best estimation $[\underline{R}]$ of the unknown Reynolds stress
  $[\mathbf{R}]$ is available at any location in the field. The estimation can be obtained in
  advance by performing a baseline RANS simulation or by incorporating observation data and
  incorporating prior corrections to the baseline RANS-predicted Reynolds stresses. Without further
  information, it is reasonable to take the best estimation $[\underline{R}]$ obtained this way as
  the mean of the true Reynolds stress $[\mathbf{R}]$, with the latter being modeled as a random
  variable.  This constraint is expressed as follows:
  \begin{equation}
    \label{eq:expect}
  \mathbb{E}\{ [\mathbf{R}] \} = [\underline{R}] .
  \end{equation}
\end{enumerate}

  The theories on the maximum distribution of random matrices developed in the earlier
  works of Soize and co-workers focused primarily on positive definite matrices, but the Reynolds
  stress tensors of concern in the present application are positive semidefinite only. Although we
  have claimed that the difference has no practical consequences, we shall still carefully
  reconciled the differences before utilizing the theories of Soize~\cite{soize2000nonparametric}.
  In order to obtain a maximum entropy distribution for a positive definite matrix that satisfies
  the normalization and mean constraints similar to those outlined above,
  Soize~\cite{soize2000nonparametric} took a two-step approach. Specifically, he first obtained the
  PDF for a \emph{normalized, positive definite} random matrix~$[\mathbf{G}]$ whose mean is the
  identity matrix $[I]$, i.e., $\mathbb{E}\{ [\mathbf{G}] \} = [I]$.  The probability distribution
  for $[\mathbf{G}]$ must also satisfy the maximum entropy condition and the normalization
  condition in Eq.~(\ref{eq:normalization}) above.  Once such a matrix $[G]$ is obtained, a matrix
  that satisfies the condition $\mathbb{E}\{ [\mathbf{R}] \} = [\underline{R}]$ can be constructed
  as follows:
\begin{equation}
  \label{eq:a-transform}
[\mathbf{R}] = [\underline{L}_R]^T  [\mathbf{G}]  [\underline{L}_R] , 
\end{equation}
where $ [\underline{L}_R]$ is an upper triangular matrix with non-negative diagonal entries obtained
from the following factorization of the specified mean $[\underline{R}]$, i.e.,
\begin{equation}
  \label{eq:lu-decomp}
  [\underline{R}] = [\underline{L}_R]^T [\underline{L}_R].
\end{equation}
In the present development, we consider the following two scenarios.  (1) When $[R] \in
\mathbb{M}_d^{+}$ is positive definite, the decomposition above is the well-known Cholesky
factorization, and the decomposition is unique. Its numerical implementations are available in many
standard numerical toolboxes (e.g., Matlab and the numpy library of Python). (2) When
  $[R] \in \mathbb{M}_d^{+0} \setminus \mathbb{M}_d^{+}$ is in the subset of singular positive
  semidefinite matrices, where $\setminus$ indicates set subtraction, it can be proved that the
decomposition in the form above still exits with $L$ having non-negative diagonal entries, although
the decomposition is no longer unique. Moreover, Cholesky factorization routines for such matrices
are still numerically stable if complete pivoting is performed~\cite{higham90analysis}. In practice,
we can make it slightly positive definite by adding a small element to the diagonal entries and thus
still use the conventional Cholesky factorization routines. 


With the reconciliation above, hereafter we only need to focus on probability distribution of the
normalized positive definite matrix $[\mathbf{G}]$ and can thus directly use the results in the
earlier works of ~\cite{soize2005random}.  By using the classical optimization techniques based on
Lagrange multipliers, the probability density function of the random matrix $[\mathbf{G}]$ is
obtained as follows~\cite{soize2005random}:
\begin{equation}
  \label{eq:pdf-g}
  p_{[\text{\textbf{\tiny G}}]}([G]) = \mathbbm{1}_{\mathbb{M}_d^+}([G])  \times C_{\text{{\tiny
        G}}} \times \det [G] \,  (d+1)(1-\delta^2)/(2 \delta^2) 
  \times \exp \left( -\frac{d+1}{2 \delta^2} \operatorname{tr} [G] \right) ,
\end{equation}
where $\mathbbm{1}_{\mathbb{M}_d^+}([G]) $ is an indicator function, i.e., it is one if $[G] \in
\mathbb{M}_d^+ $ and is zero otherwise. The positive constant $C_{\text{{\tiny G}}} $ is:
\begin{equation}
  \label{eq:cg}
  C_{\text{{\tiny G}}} = \frac{(2\pi)^{-d(d-1)/4} \left(\frac{d+1}{2\delta^2} \right)^{d(d+1)(2\delta^2)^{-1}}}
  {\prod_{j = 1}^{d} \Gamma \left( \frac{d+1}{2 \delta^2} + \frac{1-j}{2} \right) }
\end{equation}
where $\Gamma(z)$ is gamma function defined by $\Gamma(z) = \int_0^{\infty} t^{z-1} e^{-t} dt$.
The dispersion parameter $\delta$ indicates the uncertainty in the random matrix and is defined as:
\begin{equation}
  \label{eq:delta-def}
\delta = \left[ \frac{1}{d} \mathbb{E}\{ \| [G] - [I] \|_F^2 \}  \right]^{\frac{1}{2}}
\end{equation}
where $\| \cdot \|_F$ is Frobenius norm, e.g., $\| G\|_F = \sqrt{\operatorname{tr}([G]^T [G])}$.  It
can be seen that $\delta$ is analogous to the variance of a scalar random variable normalized by its
mean. It has been shown~\cite{soize2000nonparametric} that
\begin{equation}
  \label{eq:delta}
  0 < \delta < \sqrt{\frac{d+1}{d+5}},
\end{equation}
which reduces to $0 < \delta < \sqrt{2}/2$ for $d=3$. This constraint is related to the positive
definiteness of matrix $ [\mathbf{G}] $.

The individual elements in the random matrix $[\mathbf{G}]$ are correlated and the same can thus be
said for the random matrix $[\mathbf{R}]$. This correlation needs to be accounted for when
generating Monte Carlo samples, which can be achieved in a similar way as in sampling from random
processes (e.g., Gaussian processes). The sampling method is detailed below.

\subsection{Monte Carlo Sampling of Reynolds Stresses at One Point}
\label{sec:mc-1pt}
Soize~\cite{soize2000nonparametric} outlined a Monte Carlo method to sample the probability
distribution for random matrix $[\mathbf{R}]$ for a given dispersion parameter $\delta$ and mean
$[\underline{R}]$.  To proceed with the sampling, the random matrix with identity mean is first
represented by its Chelosky factorization:
\begin{equation}
    \label{eq:G}
    [\mathbf{G}] = [\mathbf{L}]^T [\mathbf{L}] ,
  \end{equation}
  where random matrix $[\mathbf{L}]$ are upper triangular matrices with independent elements. In the
  procedure, each element $\mathbf{L}_{i j}$ in matrix $[\mathbf{L}]$ is sampled independently with
  the algorithms below (see Section 5.1 in ref.~\cite{soize2005comprehensive}):
  \begin{enumerate}
  \item The off-diagonal elements $\mathbf{L}_{i j}$ with $i < j$ are obtained from
    \begin{equation}
      \label{eq:L-offdiag}
      \mathbf{L}_{i j} = \sigma_d  \; \mathbf{w}_{i j}, 
    \end{equation}
    in which $\sigma_d = \delta \times (d+1)^{-1/2}$, and $\mathbf{w}_{i j}$ (i.e.,
    $\mathbf{w}_{12}$, $\mathbf{w}_{13}$, and $\mathbf{w}_{23}$) are independent Gaussian random
    variables with zero mean and unit variance.
  \item The diagonal elements are generated as follows:
    \begin{equation}
      \label{eq:L-diag}
      \mathbf{L}_{i i} = \sigma_d \sqrt{2 \mathbf{u}_i} \quad \textrm{ with } 
      \; i = 1, 2, 3
    \end{equation}
    where $\mathbf{u}_i$ is a positive valued gamma random variable with the following
    probability density function:
    \begin{equation}
      \label{eq:v-pdf}
      p(u) = \mathbbm{1}_{\mathbb{R}^+}(u) \; \frac{u^{\frac{d+1}{2\delta^2} + \frac{1-i}{2}
          -1 } \exp(-u)}{\Gamma \left( \frac{d+1}{2\delta^2} + \frac{1-i}{2} 
        \right)} 
    \end{equation}
    That is, the $i^{\textrm{th}}$ diagonal term $\mathbf{u}_i$ conforms to gamma distribution with
    shape parameter $k = (d+1)/(2\delta^2) + (1-i)/2$ and scale parameter 1.
\item After obtaining the elements as above, a realization $[L]$ of the random matrix $[\mathbf{L}]$
  can be assembled. Subsequently, realizations $[G]$ and $[R]$ for matrices $[\mathbf{G}]$ and
  $[\mathbf{R}]$, respectively, can be obtained from $[\mathbf{G}] = [\mathbf{L}]^T [\mathbf{L}]$
  and $[\mathbf{R}] = [\underline{L}_R]^T [\mathbf{G}] [\underline{L}_R]$ as in Eqs.~(\ref{eq:G})
  and~(\ref{eq:a-transform}).
  \end{enumerate}

\subsection{Random Matrix Field Model for Reynolds Stresses}
\label{sec:mc-field}
The probability model and the Monte Carlo sampling algorithm described in Sections~\ref{sec:pdf-1pt}
and~\ref{sec:mc-1pt} above concerns the Reynolds stress tensor at a single point. An important
challenge in the present application is to represent and model the correlation of Reynolds stress at
different spatial locations. This feature distinguishes the present contribution from previous
applications of random matrix in composite materials models\cite{das2009bounded} and structural
dynamics~\cite{soize2013stochastic}.

In this work we assume that the off-diagonal terms and the square roots of the diagonal terms have
the same spatial correlation structures.  Gaussian kernel is among the most common choice in the
literature, but other kernels such as periodic (e.g., sinusoidal)
kernels~\cite{guilleminot2012stochastic} or exponential kernels~\cite{sakamoto2002polynomial} have
been used as well. The choice of kernel is reflected in the smoothness of the
  realizations of the random tensor field $\mathbf{L}$.  With the Gaussian kernel used in this
work, the correlation between two random variables at two spatial locations $x$ and $x'$ can be
written as:
\begin{subequations}
  \label{eq:Lx}
\begin{align}
  \rho_L \{ \mathbf{L}_{i j}(x) , \mathbf{L}_{i j}(x') \} & \equiv \frac{ \operatorname{Cov} \{
    \mathbf{L}_{i j}(x) , \mathbf{L}_{i j}(x') \} } {\sigma_{i j}(x) \sigma_{i j}(x')} = \exp \left[
    -\frac{|x-x'|^2}{l^2} \right]
  \quad \textrm{ for } \quad i < j  \\
  \rho_L \{ {\mathbf{L}_{i i}^2(x)} , {\mathbf{L}_{i i}^2(x')} \} & \equiv \frac{ \operatorname{Cov}
    \{ \mathbf{L}^2_{i i}(x), \mathbf{L}^2_{i i}(x') \} } {\sigma_{i i}(x) \sigma_{i i}(x')} = \exp
  \left[ -\frac{|x-x'|^2}{l^2} \right]  \quad x, x' \in \Omega
\end{align}  
\end{subequations}  where $\Omega$ is the spatial domain of the flow field.  Note
that repeated indices \emph{do not} imply summation.  $\sigma_{i j}^2(x)$ and $\sigma_{i i}^2(x)$
are the variances of the marginal distributions for the off-diagonal elements $\mathbf{L}_{i j}(x)$
and the square of the diagonal elements $\mathbf{L}^2_{i i}(x)$, respectively. That is,
$\operatorname{var} \{ \mathbf{L}_{i j}(x) \} = \sigma^2_{ij}(x)$ and $\operatorname{var} \{
\mathbf{L}^2_{i i}(x) \} = \sigma^2_{ii}(x)$, the expression of which will be given in
Eqs.~(\ref{eq:sigma-ij}) and (\ref{eq:sigma-ii}), respectively.  $\rho_L$ is the correlation kernel,
$|\cdot|$ is the Euclidean norm of vectors, and $l$ is the correlation length scale of the
stochastic process.  We emphasize that the correlations above are between \emph{the same element at
  different locations} $x$ and $x'$, and that the random variables $\mathbf{L}_{11}$,
$\mathbf{L}_{22}$, $\mathbf{L}_{33}$, $\mathbf{L}_{12}$, $\mathbf{L}_{13}$, and $\mathbf{L}_{23}$,
which are different elements of the matrix $[\mathbf{L}]$, are still independent of each other as
pointed out above.

With the correlation model defined in Eq.~(\ref{eq:Lx}) for the elements of matrix $[\mathbf{L}]$,
the random variables used to generate $L_{i j}$ in Eqs.~(\ref{eq:L-offdiag}) and~(\ref{eq:L-diag})
can be generalized to random fields. First, it is straightforward to generalize the Gaussian random
variable $\mathbf{w}_{i j}$ to a Gaussian random field (also referred to as Gaussian process),
\begin{equation}
  \label{eq:L-diag-sp}
  \mathbf{L}_{i j}(x) = \sigma_d(x) \mathbf{w}_{i j} (x) ,
\end{equation}
where $\mathbf{L}_{i j}(x) $ and $\mathbf{w}_{i j} (x) $ are random fields indexed by the spatial
coordinate $x$, and
\begin{equation}
  \label{eq:sigma-d}
\sigma_d(x) = \delta(x) \,  (d+1)^{-1/2}  
\end{equation}
is a spatially varying field specifying the standard deviation of the marginal distribution of $L_{i
  j}(x) $ at location $x$. Simple algebra shows that if we choose $\mathbf{w}_{i j} (x) \sim
\mathcal{GP}(0, K)$ as the Gaussian process with covariance kernel $K(x, x') = \exp \left[ -
  (|x-x'|^2)/l^2\right]$ as in Eq.~(\ref{eq:Lx}), then
\begin{equation}
  \label{eq:sigma-ij}
  \sigma_{ij}(x) = \sigma_d(x)  = \frac{\delta(x)}{\sqrt{d+1}} , \quad \textrm{or equivalently} \quad
  \sigma^2_{ij}(x) = \frac{\delta^2(x)}{d+1}. 
\end{equation}
The Gaussian processes $\mathbf{w}_{i j}(x)$ can be expressed by using the Karhunen--Loeve (KL)
expansion, which can be written as follows when truncated to $N_{\textrm{KL}}$ terms:
  \begin{align}
    \mathbf{w}_{i j}(x) & = \sum_{\alpha=1}^{N_{\textrm{KL}}} \; \sqrt{\lambda_\alpha} \, \phi_\alpha(x) \;
    \boldsymbol{\omega}_\alpha \label{eq:kl-reconstruct}
  \end{align}
  where $\boldsymbol{\omega}_\alpha \sim \mathbf{N}(0, 1)$ with $\alpha = 1, \cdots,
  N_{\textrm{KL}}$ being the standard Gaussian random variables; $\lambda_\alpha$ and $\phi_\alpha (x)$
  are the eigenvalue and the corresponding eigenfunction for the $\alpha$\textsuperscript{th} mode
  obtained by solving the Fredholm integral equation~\cite{le2010spectral}:
  \begin{equation}
    \label{eq:kl}
    \int_{\Omega} \rho(x, x') \phi(x') \, dx'  = \lambda \phi(x) \, ,
  \end{equation}
  where ${\Omega}$ is the spatial domain of the flow as defined above.

  \emph{Remarks on using nonstationary and anisotropic kernels to reflect the structure of the flow
    field.}  Non-stationarity and anisotropy in the Gaussian process $\mathbf{w}(x)$ can be achieved
  by using spatially varying and/or anisotropic length scale $l(x)$, which depends on the
  characteristics of the mean flow. For example, for a nonstationary Gaussian process, the length
  scale $l$ in Eq.~(\ref{eq:Lx}) can be written as $l = \sqrt{l(x) l(x')}$ following
  ref.~\cite{dow14optimal}. As such, the physical structure of the flow field is encoded in the
  Karhunen--Loeve basis functions. It is also desirable to specify different variance
  $\sigma_{ij}(x)$ at different locations. For example, in regions with flow separation and pressure
  gradient, the Boussinesq assumptions are violated and thus most RANS models have poor
  performance. Consequently, the variance $\sigma_{ij}(x)$ should be larger in these regions, which
  can be achieved by specifying the dispersion parameter $\delta(x)$ to depend on the location~$x$.

  The sampling algorithm for the random variables of the diagonal terms in Eq.~(\ref{eq:L-diag}) can
  be similarly generalized to random fields as follows:
\begin{equation}
  \label{eq:L-offdiag-sp}
  \mathbf{L}_{ii}(x) = \sigma_d(x) \sqrt{2 \, \mathbf{u}_i(x)}   ,
\end{equation}
where $\sigma_d(x)$ is defined the same as above. Recall the facts that $\mathbf{u}_i(x)$ has a
gamma distribution with shape parameter $k = (d+1)/(2\delta^2) + (1-i)/2$ and that
$\operatorname{Var}\{ \mathbf{u}_i \} = k $. Substituting Eq.~(\ref{eq:sigma-d}) into
Eq.~(\ref{eq:L-offdiag-sp}) and using the two facts above, after some algebra one can show that
\begin{equation}
  \label{eq:sigma-ii}
  \sigma^2_{ii} \equiv   \operatorname{Var}\{ \mathbf{L}^2_{ii} \} =  4 \sigma_d^4
  \operatorname{Var}\{ \mathbf{u}_i \} =
  \frac{2 \delta^2}{d+1} + \frac{2 \delta^4}{(d+1)^2} \times (1-i)  \quad 
\textrm{for} \quad i = 1, 2, 3,
\end{equation}
where $d=3$, and the spatial dependence of all other variables are omitted for
notational brevity.  However, since $\mathbf{u}_i(x)$ are random fields with gamma marginal
distributions, they are not straightforward to synthesize. Although the KL expansion above can still
be used to represent the random field $\mathbf{u}_i(x)$, the corresponding coefficients obtained
from the expansion are neither independent nor Gaussian. To overcome this difficulty, we follow the
procedure in ref.~\cite{sakamoto2002polynomial} and express the gamma random variables
$\mathbf{u}_i$ at any location $x$ with polynomial chaos expansion:  
  \begin{align}
    \mathbf{u}(x) & = \sum_{\beta = 0}^{N_p} U_\beta(x) \Psi_\beta(\mathbf{w}(x))   \label{eq:polyexp}
    \\
    &  = U_0 + U_1 \mathbf{w}(x) + U_2  (\mathbf{w}^2(x) - 1) 
    + U_3 (\mathbf{w}^3(x) - 3 \mathbf{w}(x)) 
    + U_4 (\mathbf{w}^4(x) - 6 \mathbf{w}^2(x) + 3) + \cdots  \notag
  \end{align}
  where $N_p$ is the order of polynomial expansion; for a given $x$, $\Psi_\beta(\mathbf{w}(x))$ are
  the Hermite polynomials in standard Gaussian random variable $\mathbf{w}(x)$, with the first four
  terms explicitly written as above; $U_\beta$ is the coefficients for the $\beta^{\textrm{th}}$
  polynomial, for which the spatial index $x$ has been omitted for brevity. The coefficients can be
  obtained from orthogonality conditions with respect to the Gaussian
  measure~\cite{xiu2010numerical}:
\begin{equation}
  \label{eq:U_i}
  U_\beta  = \frac{\langle \mathbf{u} {\Psi}_\beta \rangle}{\langle {\Psi}_\beta^2 \rangle} \\
   =  \frac{1}{\langle {\Psi}_\beta^2 \rangle} \int_{-\infty}^{\infty}
   F_{\mathbf{u}}^{-1}[F_{\mathbf{w}}(w)] \; {\Psi}_\beta({w})  \; p_{\text{\tiny \textbf{w}}}(w) dw
 \end{equation}
 where $\langle \boldsymbol{\Psi}_\beta^2 \rangle$ is the variance of $i$\textsuperscript{th} order
   polynomial of standard Gaussian random variable; $F_{\text{\tiny
       \textbf{w}}}(\mathbf{w})$ and $p_{\text{\tiny \textbf{w}}} (\mathbf{w})$ are the cumulative
   distribution function (CFD) and PDF, respectively, of the standard Gaussian variable;
$F_{\mathbf{u}}$ and $F_{\mathbf{u}}^{-1}$ are the CDF and its inverse, respectively, of random
variable $\mathbf{u}$.   As indicated previously, Eq.~(\ref{eq:U_i}) is evaluated for
  each $x$ with $\mathbf{w}(x)$ being a standard Gaussian variable.

Sakamoto and Ghanem~\cite{sakamoto2002polynomial} have derived the relation between spatial
correlation of the non-Gaussian field $\mathbf{u}(x)$, i.e., $\rho_u \{\mathbf{u}(x), \mathbf{u}(x')
\}$ and the correlation $\rho_w \{\mathbf{w}(x),\mathbf{w}(x')\}$ of the Gaussian random field
$\mathbf{w}(x)$ used in its polynomial chaos representation. This is achieved by substituting
Eq.~(\ref{eq:polyexp}) into the definition of covariance. They further showed that the following
approximation yields very good accuracy:
\begin{equation}
  \label{eq:rhou-rhow}
  \rho_u \{\mathbf{u}(x), \mathbf{u}(x') \} \approx \rho_w \{\mathbf{w}(x), \mathbf{w}(x')\} \, .
\end{equation} 
We have verified that the two are indeed approximately equal based on the generated samples. This
  approximation above is what we shall use in our work.
Therefore, given the correlation structure of field $\mathbf{u}_i$, an approximation of the kernel
$\rho_w \{\mathbf{w}(x), \mathbf{w}(x')\}$ is obtained, which can be used in the KL expansion to
find the eigenmodes and further synthesize the Gaussian random fields $\mathbf{w}_{ii}$. Finally,
the non-Gaussian random field $\mathbf{u}_i(x)$ can be obtained by reconstructing from the
polynomial chaos based on Eq.~(\ref{eq:polyexp}) at each location  $x \in \Omega$.

\subsection{Monte Carlo Sampling of Reynolds Stress Fields}
\label{sec:algo-summary}

Based on the development above, the algorithm of Monte Carlo sampling of Reynolds stress field can
be performed as follows:

\begin{enumerate}
\item Draw independent samples $\omega_\alpha$ from the standard Gaussian distribution
  $\mathbf{N}(0, 1)$ with $\alpha = 1, \cdots, N_{\textrm{KL}}$.
\item Use the KL expansion in Eq.~(\ref{eq:kl-reconstruct}) to reconstruct a Gaussian random field
  $w(x)$ with $\lambda_\alpha$ and $\phi_\alpha(x)$ obtained from solving Eq.~ (\ref{eq:kl}).
\item Repeat steps 1--2 above six times to obtain six independent realizations of Gaussian random
  fields $w_{12}(x)$, $w_{13}(x)$, $w_{23}(x)$, $w_{11}(x)$, $w_{22}(x)$, and $w_{33}(x)$. The first three
  will be used to synthesize the off-diagonal elements $L_{12}(x)$, $L_{13}(x)$, and $L_{23}(x)$,
  respectively, and the last three for the diagonal elements $L_{11}(x)$, $L_{22}(x)$, and $L_{33}(x)$,
  respectively.

   \emph{Remark on the choice of KL modes for the representation of the six Gaussian
      random fields.} In this work the same KL modes $\phi_{\alpha}$ and eigenvalues
    $\lambda_{\alpha}$ are used for generating all six realizations $w(x)$ of Gaussian random fields
    above.  This is a modeling choice made for simplicity and also for the lack of prior knowledge
    to justify otherwise. However, in cases where it is desirable to use a different number
    $N_{\textrm{KL}}$ of modes or even a different kernel $K(x, x')$ for each field, it can be
    achieved in a straightforward manner as well.

\item Synthesize off-diagonal elements based on Eq.~(\ref{eq:L-diag-sp}).
\item Reconstruct the random fields $\mathbf{u}_1(x)$, $\mathbf{u}_2(x)$, and $\mathbf{u}_3(x)$
  based on the polynomial chaos expansion in Eq.~(\ref{eq:polyexp}) and further obtain the fields
  $\mathbf{L}_{11}(x)$, $\mathbf{L}_{22}(x)$, and $\mathbf{L}_{33}(x)$ based on
  Eq.~(\ref{eq:L-offdiag-sp}).
\end{enumerate}

\subsection{Considerations in Numerical Implementation}
\label{sec:implement}

For clarity we outline the complete algorithm of sampling of Reynolds stress fields and
the propagation to velocities and other QoIs in \ref{sec:summary}. Remarks are made below on
computational considerations in implementing the proposed algorithm.

In the expansions described in Sections~\ref{sec:mc-1pt} and~\ref{sec:mc-field}, the Cholesky
factorization of the Reynolds stress, the polynomial chaos expansion of the marginal distribution
of the gamma random variable $\mathbf{u}_i$, and the KL expansion of the Gaussian random field all
need to be performed on a mesh.  Theoretically, they all can be performed on the RANS simulation
mesh. However, this choice would lead to unnecessarily high computational costs. The RANS mesh is
designed to resolve the flow field features. Usually it is refined in the near-wall region to
resolve the boundary layer. On the other hand, the mesh for the KL expansion only needs to resolve
the length scale of the correlation, i.e., the field $l(x)$. Finally, the mesh needed for the
polynomial chaos expansion need to resolve the variation of the dispersion parameter $\delta(x)$
field. If the dispersion is constant throughout the domain, one polynomial chaos expansion is
sufficient. Therefore, the three expansions and the corresponding reconstructions below are
performed on separate meshes, which are referred to as RANS mesh, PCE mesh, and KL mesh,
respectively. Interpolations are needed when performing operations on fields based on different
meshes. For example, in Eq.~(\ref{eq:polyexp}) where the gamma field is reconstructed from the
coefficients $U_\beta(x)$ and the Hermite polynomials in the Gaussian random field $\mathbf{w}(x)$,
the coefficients are based on the PCE mesh while the Gaussian random field is based on the KL
mesh. An interpolation of the coefficients $U_\beta(x)$ from the PCE mesh to the KL mesh is thus
needed.  The obtained gamma fields $\mathbf{u}_i$, the diagonal terms $\mathbf{L}_{i i}(x)$, and the
matrix $\mathbf{G}(x)$ are thus based on the KL mesh as well. Similarly, therefore, when the
Reynolds stress is reconstructed from $[\mathbf{R}] = [\underline{L}_R]^T [\mathbf{G}]
[\underline{L}_R]$, the matrix $[\mathbf{G}]$ is first interpolated from the KL mesh to the RANS
mesh, so that the obtained Reynolds stress realization $[R]$ can be used in the RANS simulation.

\section{Numerical Results}

\begin{figure}[!htbp]
  \centering
  \includegraphics[width=0.7\textwidth]{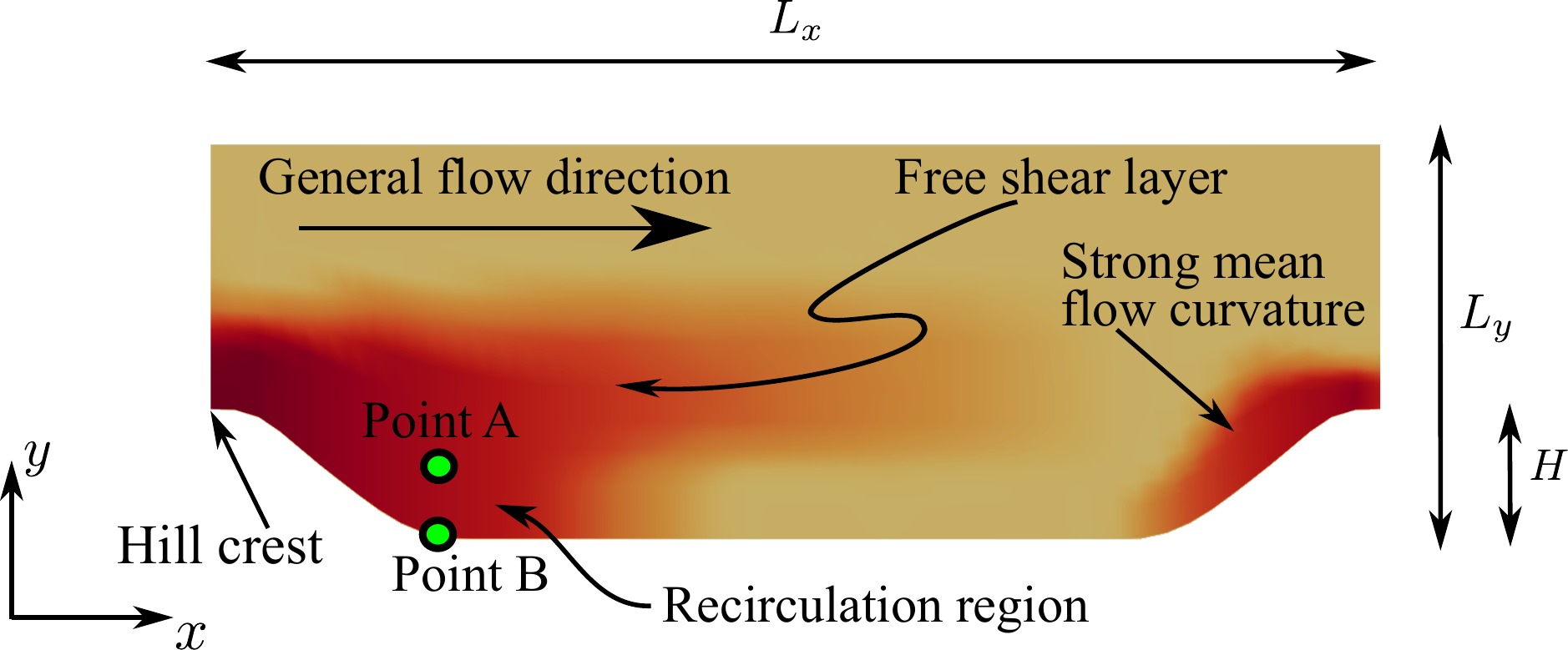}
  \caption{Domain shape for the flow in the channel with periodic hills.  The $x$-, $y$- and
    $z$-coordinates are aligned with the streamwise, wall-normal, and spanwise directions,
    respectively. All dimensions are normalized by $H$ with $L_x/H=9$ and $L_y/H=3.036$. The contour
    shows the dispersion parameter $\delta(x)$, where darker color denotes larger variance. Two typical points
    A and B are annotated in the figure.}
  \label{fig:domain_pehill}
\end{figure}

In this section we use the flow over periodic hills at Reynolds number $Re = 2800$ to demonstrate
the proposed model-form uncertainty quantification scheme for RANS simulations.  The computational
domain is shown in Fig.~\ref{fig:domain_pehill}, where all dimensions are normalized with the crest
height $H$. The Reynolds number $Re$ is based on the crest height $H$ and bulk velocity $U_b$ at the
crest.  Periodic boundary conditions are applied at the boundaries in the streamwise ($x$) direction
while non-slip boundary conditions are imposed at the walls. The spanwise direction is not
considered since the mean flow is two-dimensional. Benchmark data from direct numerical simulations
are used to compare with the sampled Reynolds stresses and the velocities obtained by propagating
the Reynolds stresses through the RANS solver.

\subsection{Cases Setup}
\label{sec:setup}

The dispersion parameter is a critical free parameter in the uncertainty quantification scheme. The
specification of $\delta$ is constrained by requirement $0 < \delta < \sqrt{2}/2$ obtained from
Eq.~(\ref{eq:delta}) and guided by the subjective belief of the user on the uncertainty in the
Reynolds stresses, i.e., larger $\delta$ indicates larger uncertainties. We investigate three cases
with different choices of dispersion parameters: (1) a small dispersion parameter $\delta = 0.2$,
(2) a large dispersion parameter $\delta = 0.6$, and (3) a spatially varying $\delta(x)$ field as
specified by the color (or grey scale) contour shown Fig.~\ref{fig:domain_pehill}.  A spatially
varying $\delta(x)$ allows the user to encode empirical knowledge about the turbulence model
and/or the flow of concern into the prior distribution of Reynolds stresses.  For example, in the
flow over periodic hills studied here, the underlying Boussinesq assumption is violated in the
regions with recirculation, non-parallel free-shear, or strong mean flow curvature (see annotations
in Fig.~\ref{fig:domain_pehill}). Consequently, these regions are particularly problematic for eddy
viscosity models (including the $k$--$\varepsilon$ model used in this work). Hence, relatively large
dispersion parameters are specified in these regions compared to other regions to reflect the lack
of confidence on the model. This is similar to the practice of specifying spatially varying variance
fields when injecting uncertainties to the physics-based quantities (magnitude and shape of the
Reynolds stress) in Xiao et al.~\cite{xiao-mfu}. On the other hand, it is also possible to use
maximum likelihood estimation to obtain a dispersion parameter $\delta(x)$ field if some measurement
data of Reynolds stresses are available, but this approach is not explored in the present work.

For Cases 1 and 2 with constant dispersion parameters in the domain, the marginal distribution of
random matrix $[\mathbf{G}]$ is the same at all locations. Therefore, the polynomial chaos expansion
needs to be performed only once. That is, the expansion coefficients $U_\beta$ in Eq.~(\ref{eq:U_i})
is spatially uniform. Hence, a PCE mesh is not needed. For Case 3 with a spatially varying
$\delta(x)$, a coarse PCE mesh of $15 \times 10$ cells is used. In all three cases, the marginal
distribution of the diagonal terms are only weakly non-Gaussian, and thus a third order polynomial
expansion is found to be sufficient.  In the KL expansion, we used 30 modes, allowing 90\% of the
variance to be captured.  Since the RANS mesh is rather small with only 3000 cells, the KL expansion
is performed on the same mesh.  In the correlation kernel we used an anisotropic yet spatially
uniform length scale ($l_x/H = 2$ and $l_y/H =1$) to reflect the aniostropy of the flow.  The
specific numbers are chosen based on the approximate length scale of the mean flow. We use 1000
samples to adequately represent the prior distribution of the Reynolds stresses. To limit the
computational cost of the RANS simulations, only 100 of the 1000 samples are randomly selected to
propagate to the velocities.  The computational parameters discussed above are summarized in
Table~\ref{tab:para}.

\begin{table}[htbp]
  \centering
  \caption{Mesh and computational parameters used in the flow over periodic hills.
    \label{tab:para}
  }
    \begin{tabular}[b]{c|c |c |c}
      \hline
      \textbf{Parameters} & \textbf{Case 1} & \textbf{Case 2} & \textbf{Case 3} \\
      \hline
      dispersion parameter $\delta(x)$ & 0.2 &  0.6 &   $0.16$--$0.65$\textsuperscript{(a)}\\
      \hline
      nature of $\delta(x)$ & uniform & uniform & nonuniform \\
      \hline
      PCE mesh & --  & -- & $15 \times 10$\\ 
      \hline
      order of  polynomial expansion ($N_p$) & \multicolumn{3}{c}{$3$}\\
      KL mesh & \multicolumn{3}{c}{$50 \times 30$}\\
      RANS mesh & \multicolumn{3}{c}{$50 \times 30$}\\
      number of KL modes  $N_{\textrm{KL}}$  & \multicolumn{3}{c}{30}\\
      correlation length scales\textsuperscript{(b)}   & \multicolumn{3}{c}{ $l_x/H = 2$, $l_y/H = 1$}\\ 
      number of Reynolds stress samples & \multicolumn{3}{c}{1000} \\
      number of   velocity propagation samples & \multicolumn{3}{c}{100}\\
      \hline            
    \end{tabular} \\
  \flushleft
    (a) Spatial contour of $\delta(x)$  is shown in Fig.~\ref{fig:domain_pehill}. \\
    (b) See Eq.~(\ref{eq:kl-reconstruct}) \\
\end{table}

\subsection{Results and Interpretations}
\label{fig:results}

As discussed above, 1000 samples of Reynolds stress random field $[\mathbf{R}](x)$ are drawn by
using the algorithm described in Section~\ref{sec:algo-summary}.  An additional challenge is the
visualization, validation, and interpretation of the sampled Reynolds stress fields, because in
general it is not straightforward to interpret the state of the turbulence from their individual
components. To facilitate physical interpretation, in all three cases we map the sampled Reynolds
stresses to the six independent physical dimensions with the scheme described in
Section~\ref{sec:physics-proj}, i.e., the magnitude $k$ of Reynolds stress tensor, its shape
(expressed in Barycentric coordinates $C_1$, $C_2$, and $C_3$), and the Euler angles ($\varphi_1$,
$\varphi_2$, and $\varphi_3$) of its eigenvectors.  The scattering of the samples in the Barycentric
triangle can be obtained from their Barycentric coordinates, and the probability density functions
for all the quantities can be estimated from the samples with a given kernel.

\begin{figure}[htbp]
  \centering 
  \subfloat[Sample scatter plot, Case 1]{\includegraphics[width=0.48\textwidth]{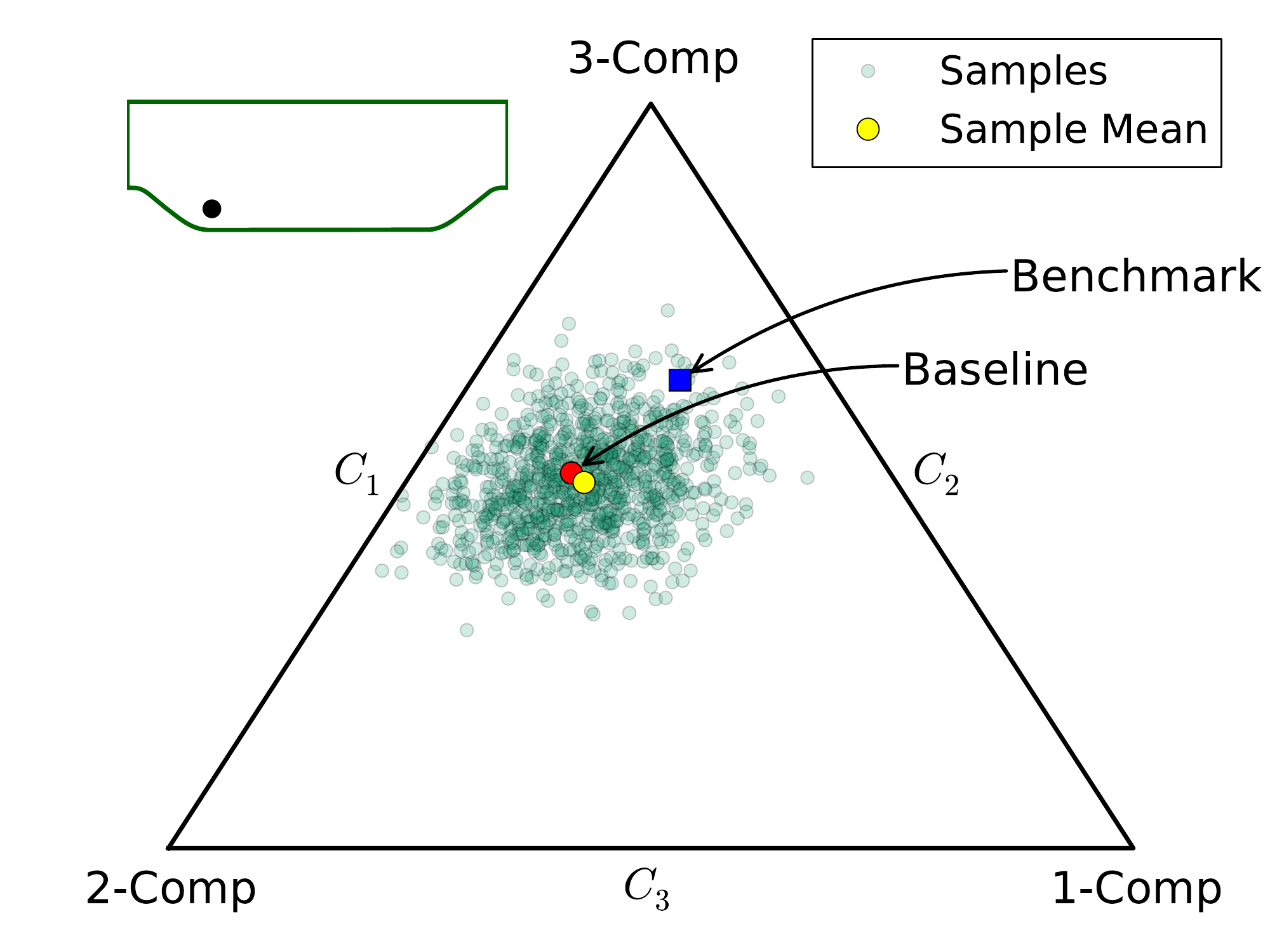}} 
  \subfloat[Sample scatter plot, Case 2]{\includegraphics[width=0.48\textwidth]{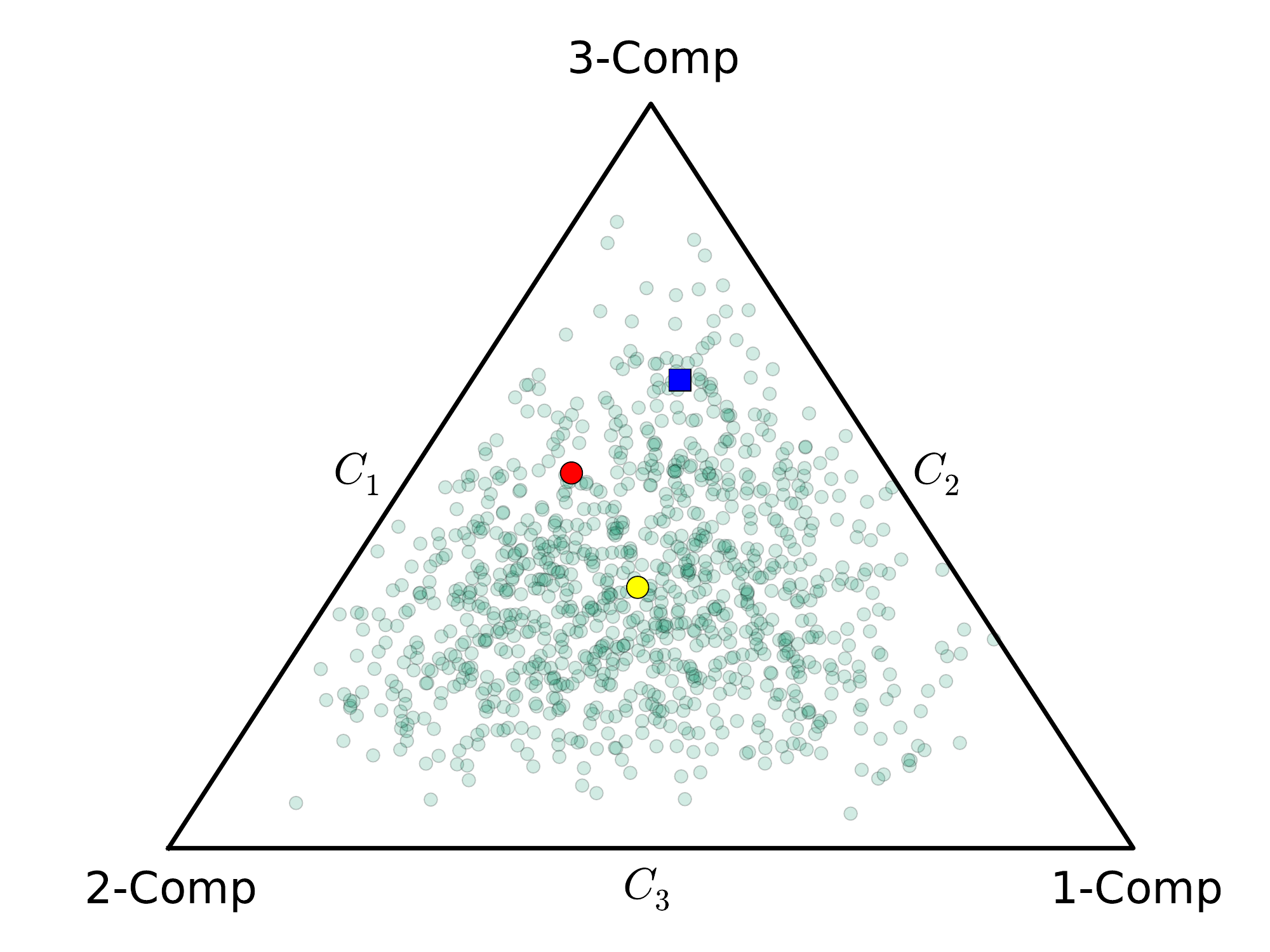}} \\
  \subfloat[Probability density contour, Case 1]{\includegraphics[width=0.48\textwidth]{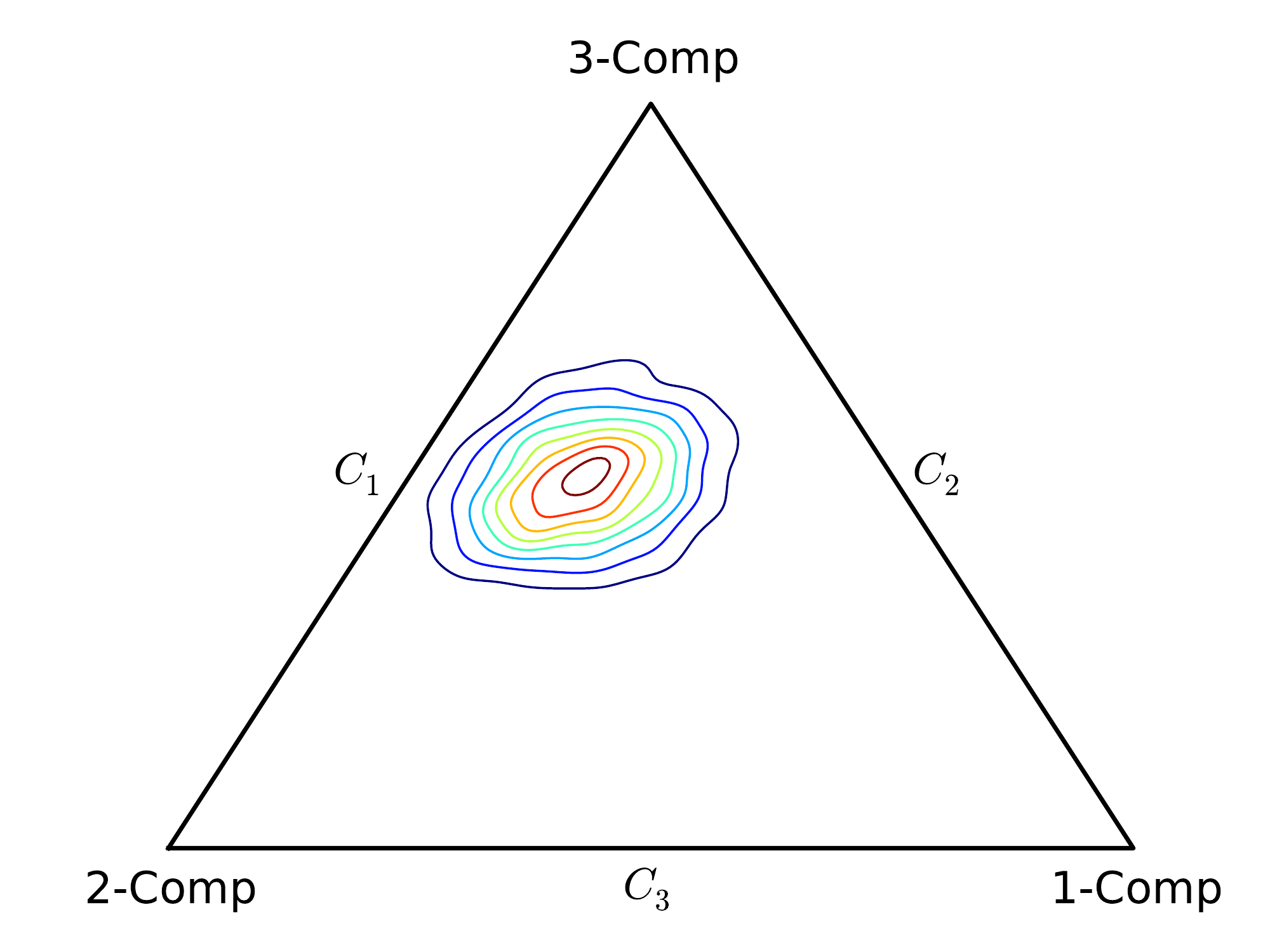}}
  \subfloat[Probability density contour, Case 2]{\includegraphics[width=0.48\textwidth]{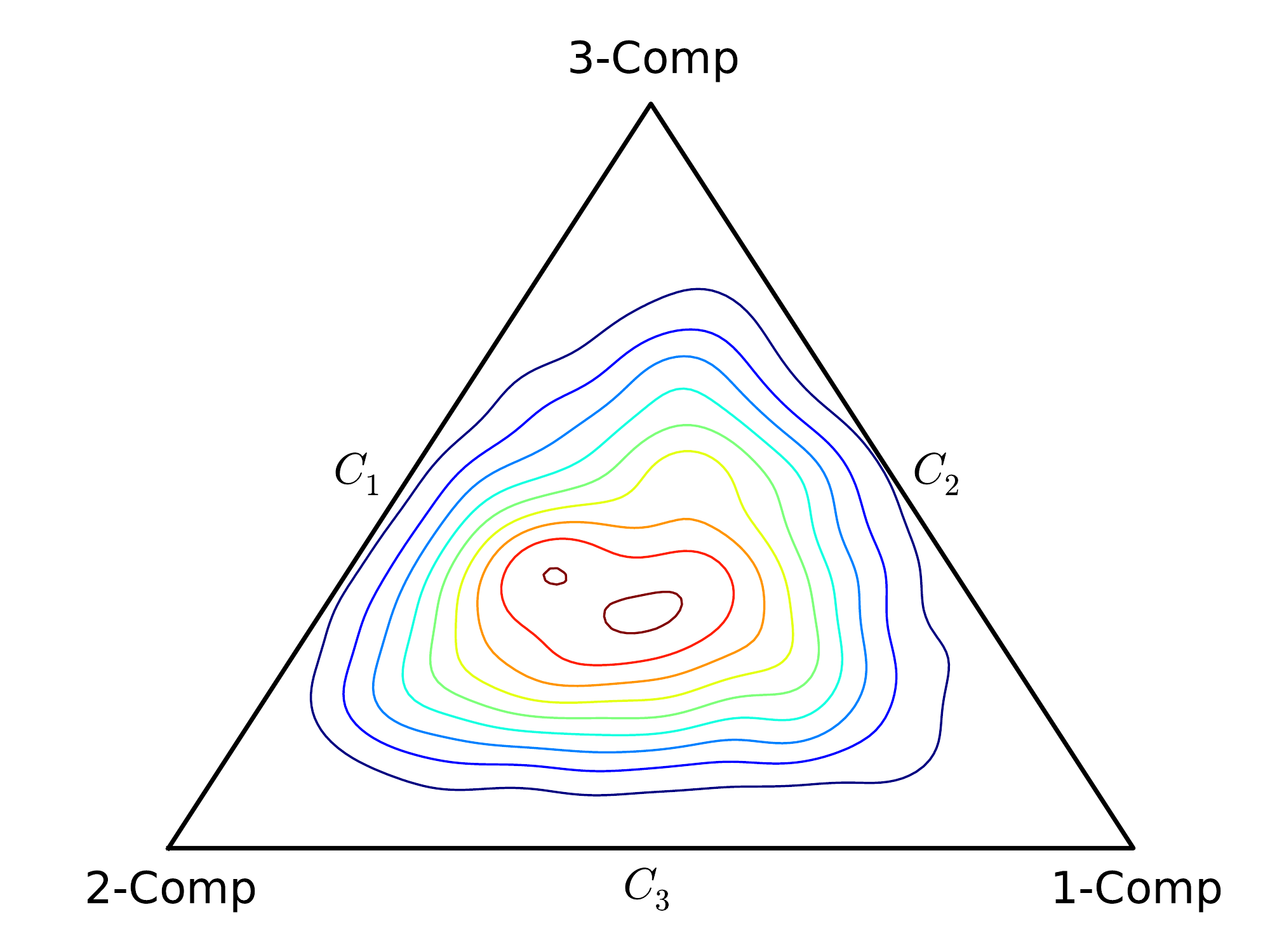}}
  \caption{Scatter plots (panels a and b) and probability density contours (panels c and d) of the
    Reynolds stress samples projected to the Barycentric coordinates for point A ($x/H=2.0,
    y/H=0.5$) located in the recirculation region. Case 1 ($\delta = 0.2$) and Case 2 ($\delta =
    0.6$) are compared.  }
  \label{fig:bayLocA}
  \end{figure}

\begin{figure}[htbp]
  \centering
   \subfloat[Scatter plot of samples]{\includegraphics[width=0.48\textwidth]{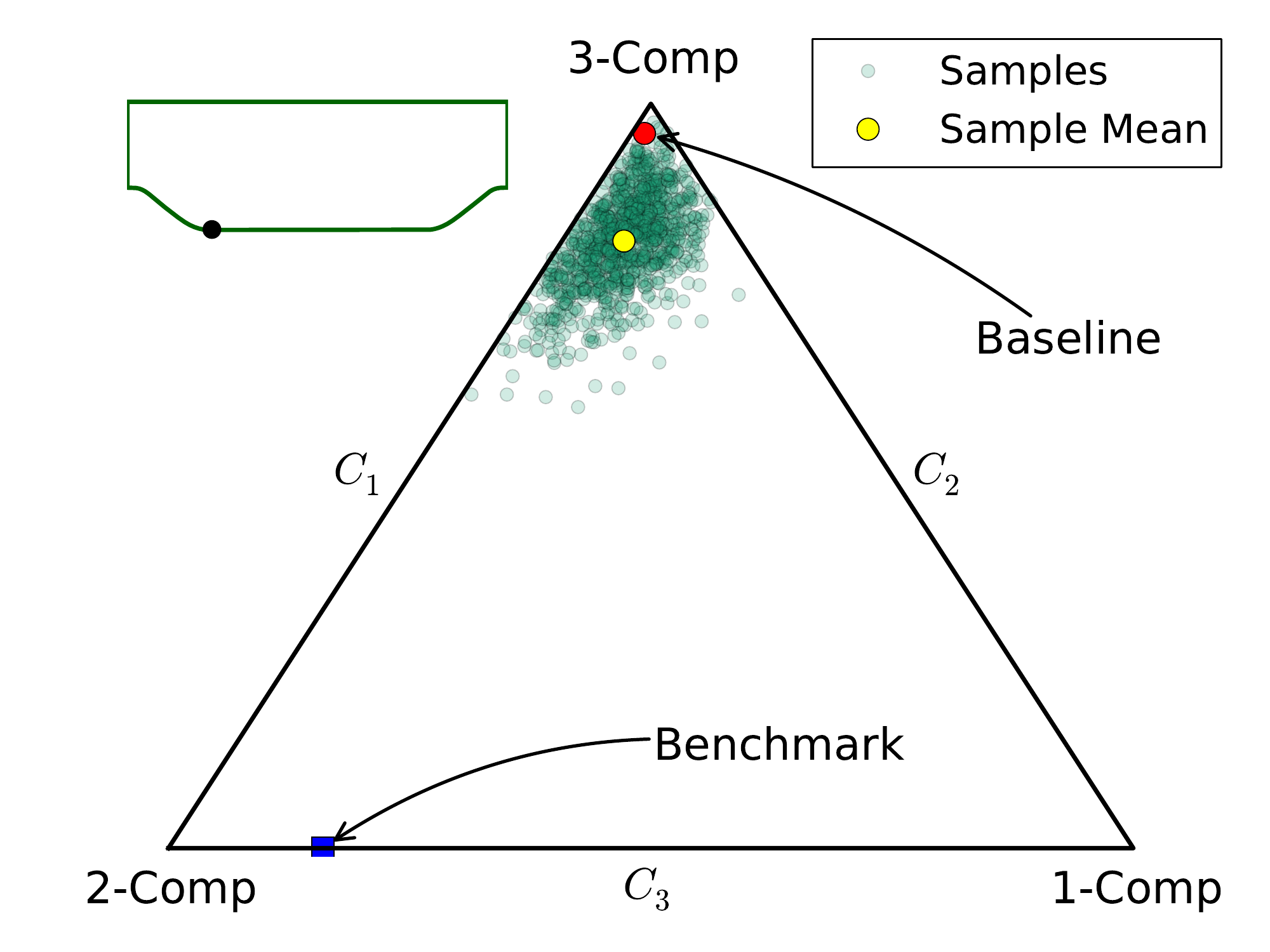}}
   \subfloat[Scatter plot of samples]{\includegraphics[width=0.48\textwidth]{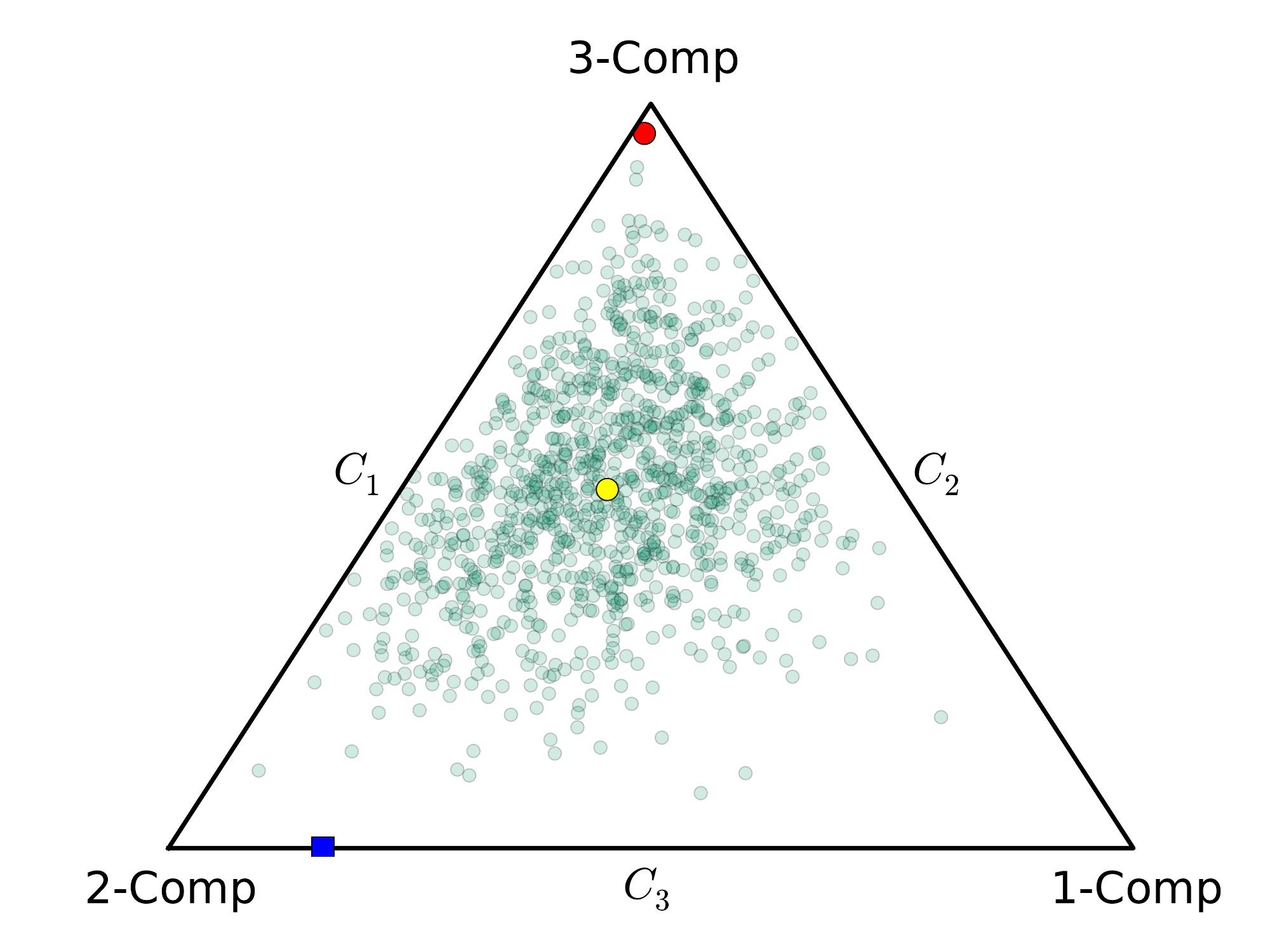}} \\
   \subfloat[Contour plot of samples]{\includegraphics[width=0.48\textwidth]{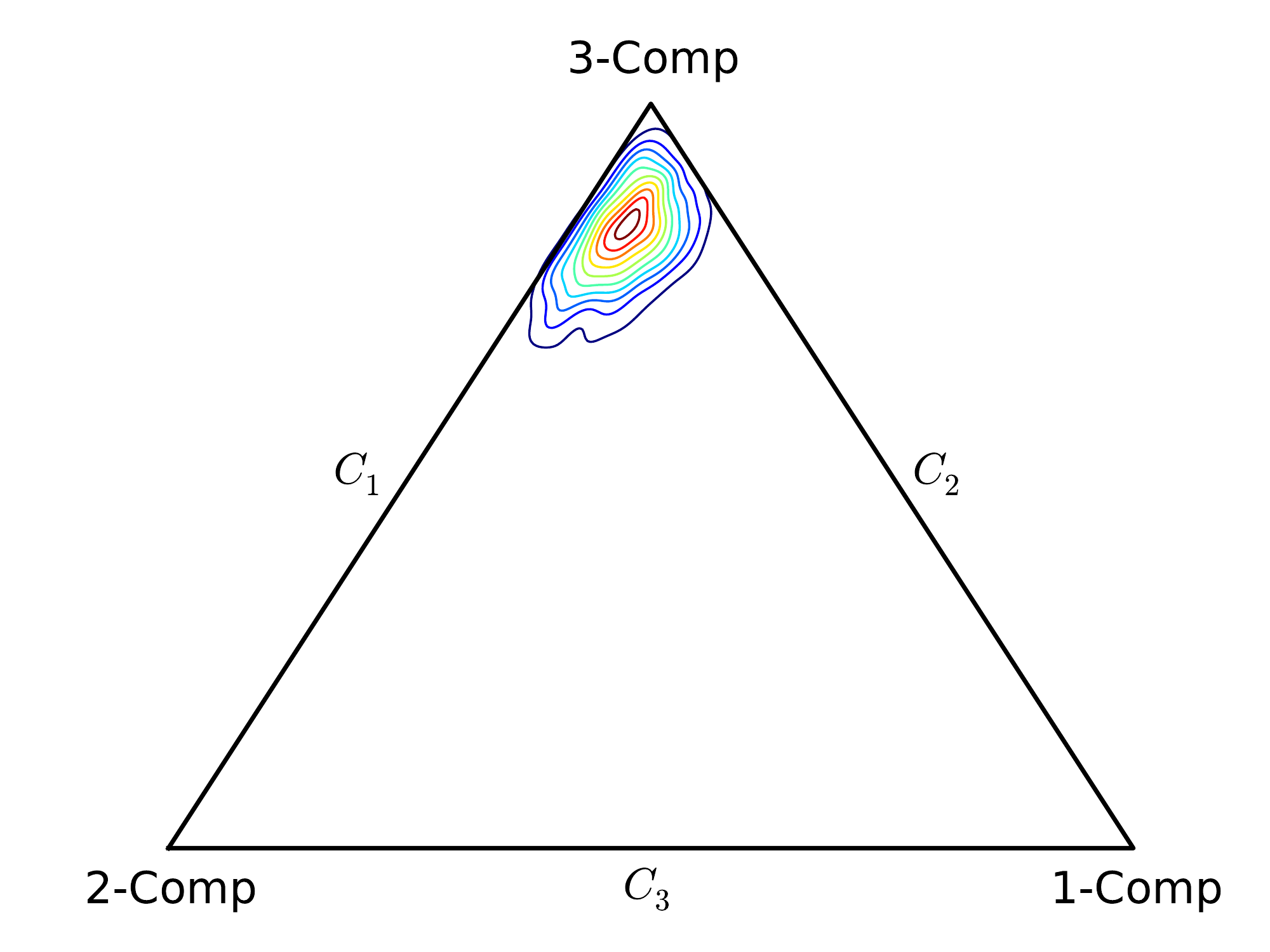}} 
   \subfloat[Contour plot of samples] {\includegraphics[width=0.48\textwidth]{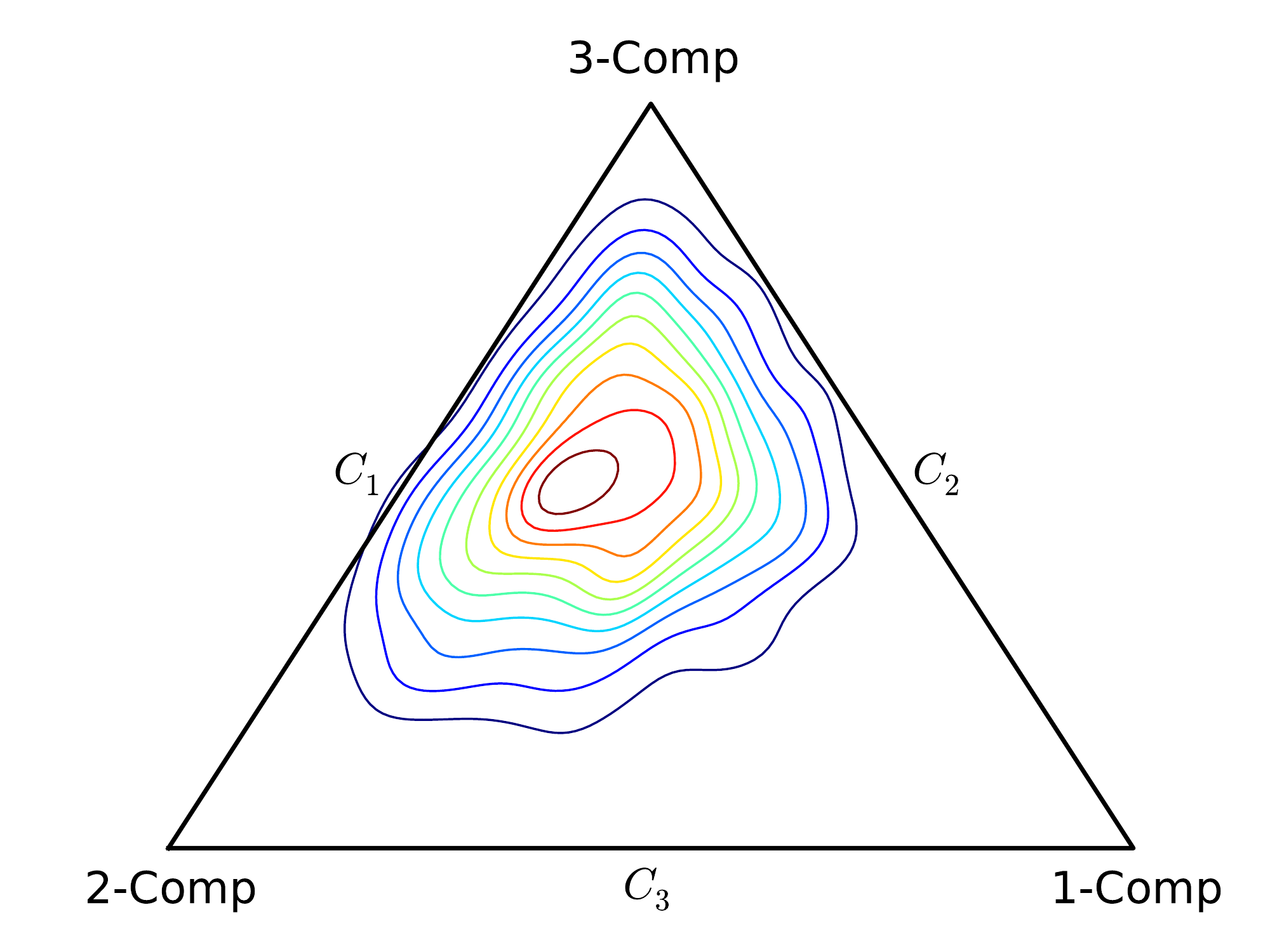}}
   \caption{Scatter plots (panels a and b) and probability density contours (panels c and d) of the
     Reynolds stress samples projected to the Barycentric coordinates for point B ($x/H=2.0,
     y/H=0.01$) located in the recirculation region. Case 1 ($\delta = 0.2$) and Case 2 ($\delta =
     0.6$) are compared. }
  \label{fig:bayLocB}
\end{figure}

The samples obtained in Case 1 (with $\delta = 0.2$) and Case 2 (with $\delta = 0.6$) are first
analyzed by visualizing the scatter plots in the Barycentric triangle, the associated probability
density contours, and the marginal distributions of the physical variables (e.g., $C_3$, $k$, and
$\varphi_3$) for two typical points: (1) point A located at $x/H = 2.0$ and $y/H = 0.5$ and (2)
point B located at $x/H = 2.0$ and $y/H = 0.01$. Point A is a generic point in the recirculation
region, and point B is a near-wall point with two-dimensional turbulence representing limiting
states.  The locations of the two points in the domain are indicated in Fig.~\ref{fig:domain_pehill}
and in the insets of Figs.~\ref{fig:bayLocA} and~\ref{fig:bayLocB}.

The scatter plots for the Reynolds stress samples at point A are shown in Figs.~\ref{fig:bayLocA}a
and ~\ref{fig:bayLocA}b for Cases 1 and~2, respectively. Since this is a generic point, the Reynolds
stress state predicted in the baseline simulation is projected to the interior of the Barycentric
triangle.  While the truth state given by the benchmark simulation is also located in the interior
of the triangle, it is further towards the bottom edge, i.e., the two-component limit (see
Fig.~\ref{fig:bary}a).  Note that the points located on the edges and vertices correspond to
limiting states as illustrated in Fig.~\ref{fig:bary}a.  In both cases the samples are scattered
around the baseline state, suggesting that perturbations introduced in the set $\mathbb{M}_d^{+0}$
about the baseline are still scattered around the corresponding baseline results in the Barycentric
triangle.  However, the two cases exhibit two notable differences. First, the scattering in Case 2
is much larger than that in Case 1, which is a direct consequence of the larger dispersion parameter
($\delta = 0.6$) compared to that in Case 1 ($\delta = 0.2$). Second, the sample mean deviates
significantly from the baseline in Case 2 while they agree quite well in Case 1. This is an
interesting observation. Recall that the probability distribution in Eq.~(\ref{eq:maxent}) is
derived under the constraint $\mathbb{E}\{ [\mathbf{R}] \} = [\underline{R}]$ given in
Eq.~(\ref{eq:expect}), which implies that the mean of the sample Reynolds stresses should be the
baseline prediction assuming the number of samples is large enough.  We have verified that the
constraint is indeed satisfied and that the sampling error is negligible with the chosen sample
number 1000.  Therefore, we can see that the mean is not preserved during the projection from the
Reynolds stress space to the Barycentric coordinates. This is due to the constraints imposed by the
positive definiteness of the Reynolds stress tensor, which manifest itself as the edge boundaries of
the Barycentric triangle. When the perturbation is small relative to the distance of the baseline
state to the boundaries, as is the situation in Case 1, the perturbation does not ``feel'' the
constraints. Consequently, the sample mean and the baseline result are approximately the same. In
contrast, when the perturbation is large compared to the distance between the baseline and one of
the boundaries, the constraint comes into play, leading to appreciable deviations between the sample
mean and the baseline. This is better illustrated with the probability density contours in
Figs.~\ref{fig:bayLocA}c and~\ref{fig:bayLocA}d for Cases 1 and~2, respectively. The contours in
Fig.~\ref{fig:bayLocA}c are approximately elliptic showing little influence from the boundaries. On
the other hand, the contours in Figs.~\ref{fig:bayLocA}d clearly follow the three edges at the
boundary and displays a triangular shape overall.  Since no samples fall outside the Barycentric
triangle as can be seen from the scatter plots in both cases, the observation also verified the
claim that the random matrix based approach proposed in this work is capable of ensuring Reynolds
stress realizability.

\begin{figure}[htbp]
  \centering
   \subfloat[Distribution  of $\Delta C_3$, Case 1]{\includegraphics[width=0.45\textwidth]{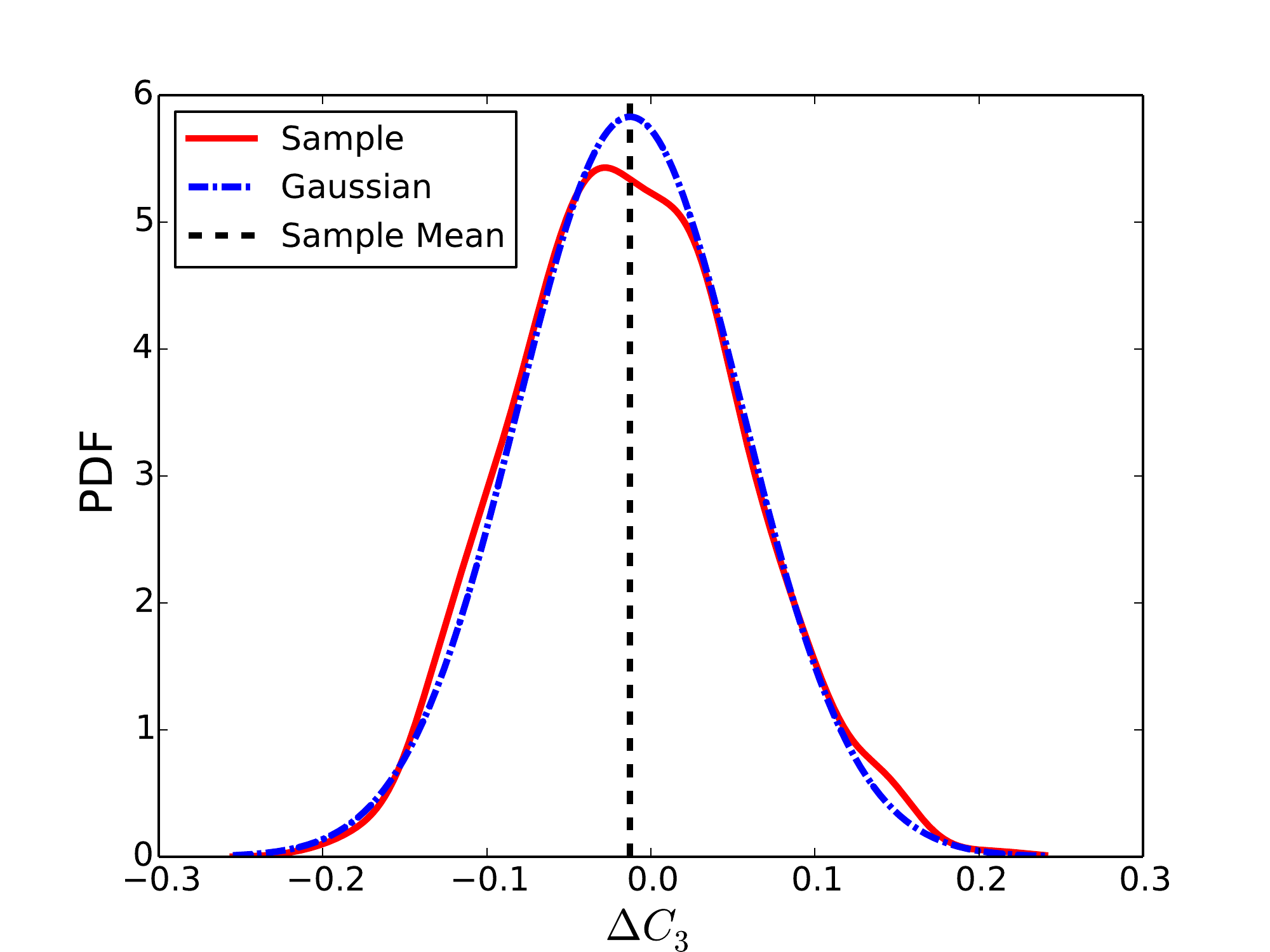}}
  \subfloat[Distribution  of $\Delta C_3$ for Case 2]{\includegraphics[width=0.45\textwidth]{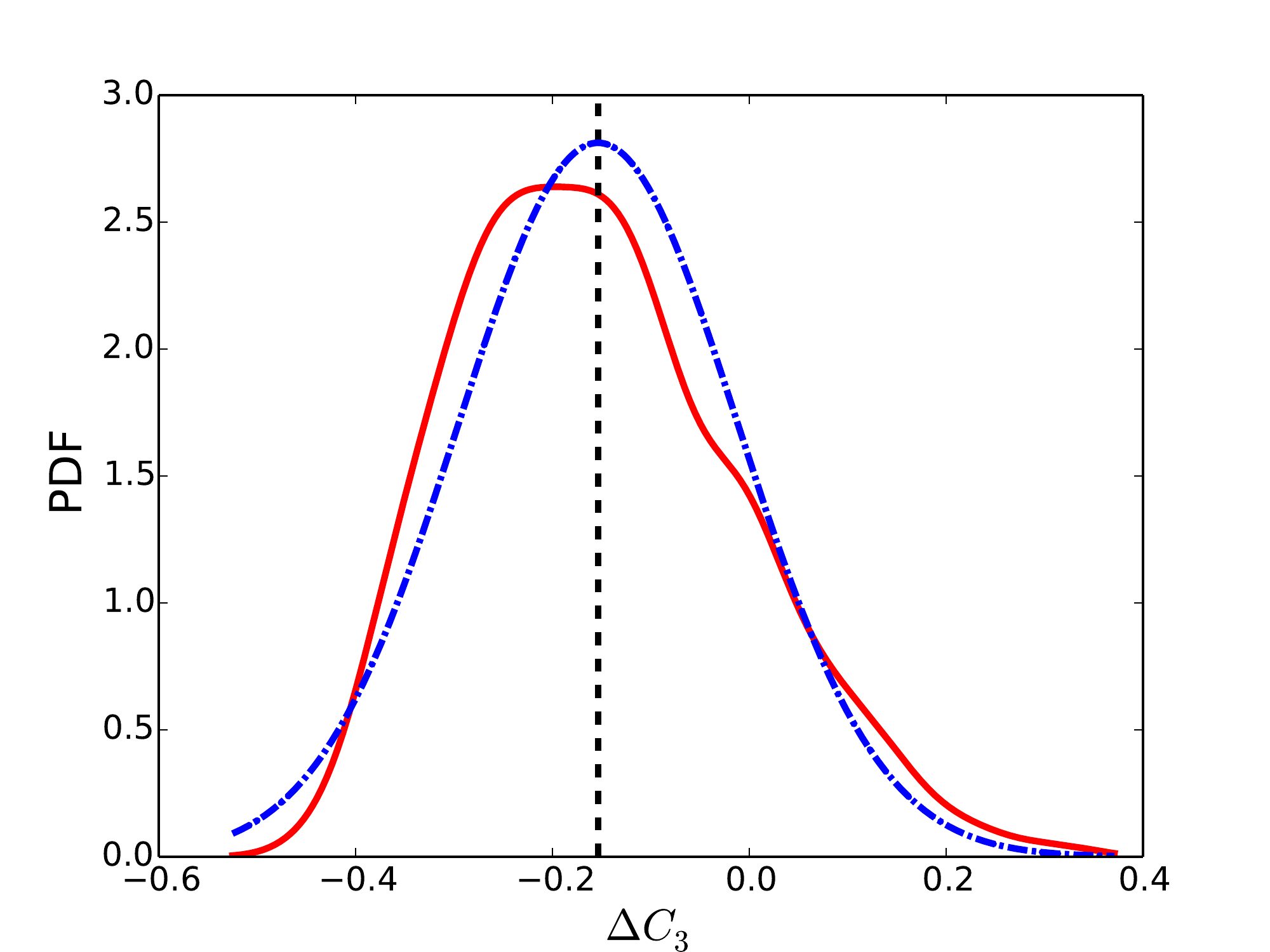}} \\
   \subfloat[Distribution  of $\Delta \varphi_3$ for Case 1]{\includegraphics[width=0.45\textwidth]{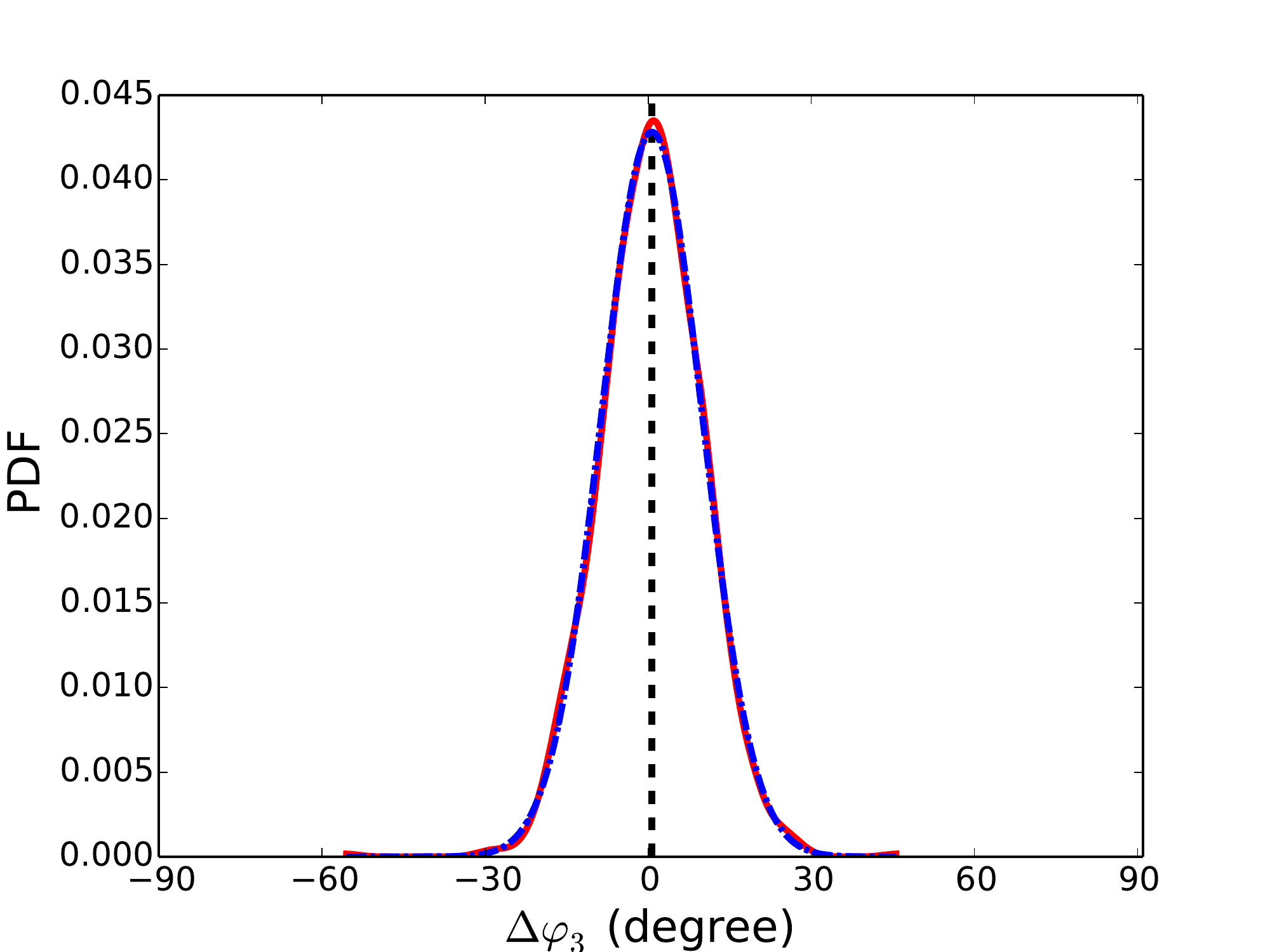}}
   \subfloat[Distribution  of $\Delta \varphi_3$ for Case 2]{\includegraphics[width=0.45\textwidth]{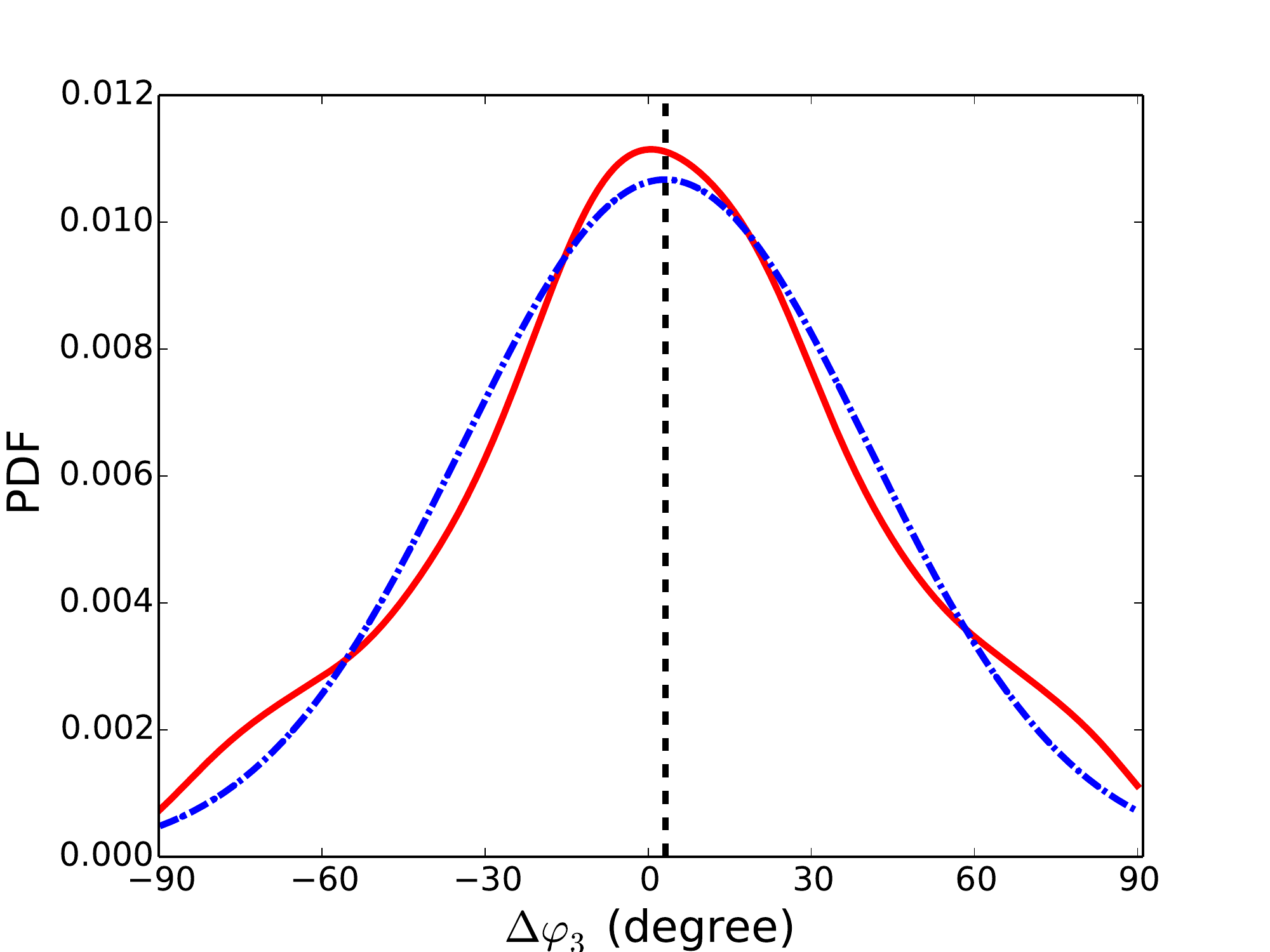}} \\
   \subfloat[Distribution $\Delta \ln k$ for Case 1]{\includegraphics[width=0.45\textwidth]{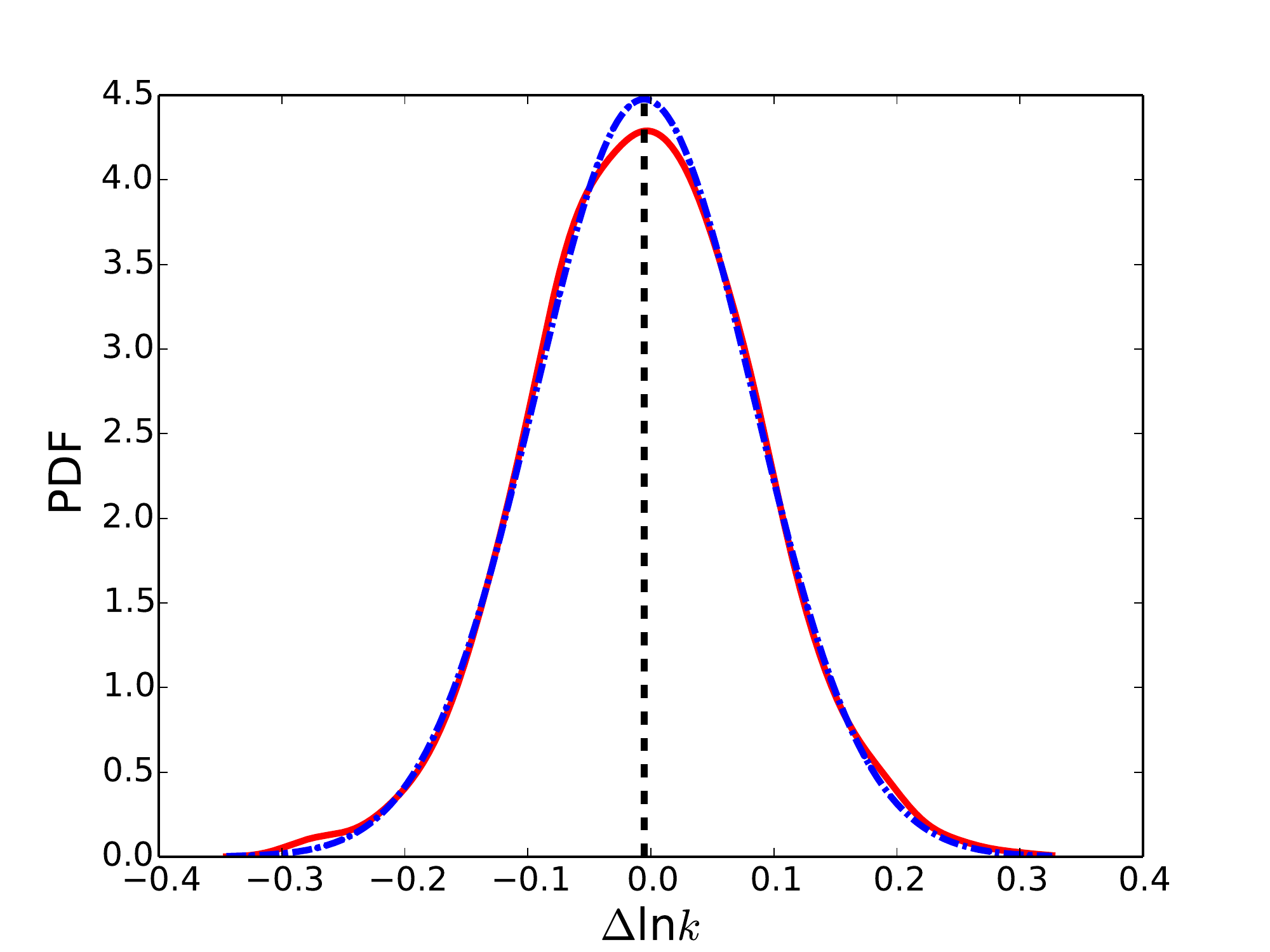}}
   \subfloat[Distribution $\Delta \ln k$ for Case 2]{\includegraphics[width=0.45\textwidth]{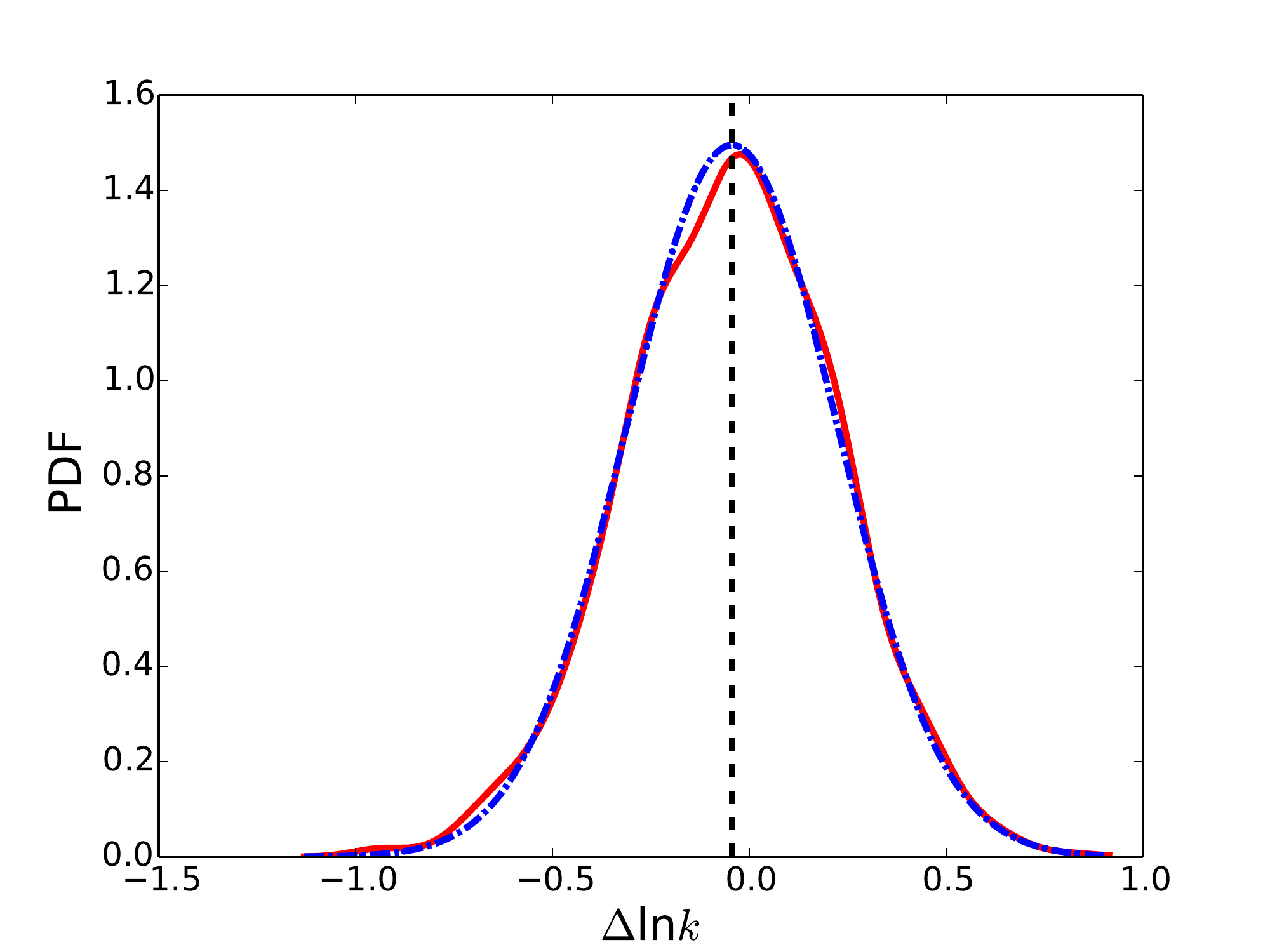}}      
   \caption{Distributions of the perturbations ($\Delta C_3$, $\Delta \varphi_3$, and
     $\Delta \ln k$ ) in the physical variables for point A ($x/H=2.0, y/H=0.5$) located in the
     recirculation region.  The distributions for Case 1 ($\delta = 0.2$) and Case 2 ($\delta =
     0.6$) are compared. }
  \label{fig:shapeLocA}
  \end{figure}

  To further examine the probability distribution of the Reynolds stress samples, in
  Fig.~\ref{fig:shapeLocA} we present the discrepancies $\Delta C_3$, $\Delta \varphi_3$, $\Delta
  \ln k$ between the sampled and the baseline Reynolds stress for point A.  Perturbations on the
  other physical quantities (e.g., the other Barycentric coordinates $\Delta C_1$ and $\Delta C_2$)
  have similar characteristics to $\Delta C_3$ presented here and thus are omitted for brevity.
  Similarly, perturbations $\Delta \varphi_1$ and $\Delta \varphi_2$ for the Euler angles
  $\varphi_1$ and $\varphi_2$ are omitted as well. Xiao et al.~\cite{xiao-mfu} introduced
  uncertainties to the predicted Reynolds stresses by modeling the discrepancies in shape and
  magnitude of the Reynolds stress as Gaussian random fields. It is illustrative to compared the
  discrepancies above (which are obtained from samples directly drawn in $\mathbb{M}_d^{+0}$) to the
  corresponding Gaussian distributions having the same mean and variance as those of the samples.
  The comparison is performed in Fig.~\ref{fig:shapeLocA}, where the PDFs of the Gaussian
  distributions obtained in this way are presented along with those estimated from samples. It can be
  seen that in Case 1 the distributions for all three discrepancy quantities $\Delta C_3$, $\Delta
  \varphi_3$, $\Delta \ln k$ are rather close to the corresponding Gaussian distributions. In
  contrast, the corresponding distributions in Case 2 deviate significantly from the Gaussian
  distributions.  It is also noted that the sample mean (denoted in vertical dashed lines; same
  hereafter) of $\Delta C_3$ deviate slightly from zero in Case 1 but significantly from zero in
  Case 2, which is consistent with the earlier observations from Figs.~\ref{fig:bayLocA}a
  and~\ref{fig:bayLocA}b.  However, the sample means for $\Delta \varphi_3$ and $\Delta \ln k$ are
  zero for both Case 1 and Case 2, despite the fact that the sample distribution are non-Gaussian for
  Case 2. This can be explained by the fact that there are no physical constraints on $\ln k$ or
  $\varphi$ (except for the range $[-\pi, \pi]$ specified in the definition). This is in contrast to
  $C_3$, which is bounded in $[0, 1]$.  Notably, Figs.~\ref{fig:shapeLocA}e and~\ref{fig:shapeLocA}f
  seem to suggest that $k$ (which is the turbulent kinetic energy defined as $\frac{1}{2}
  \operatorname{tr}([R])$) approximately follows a log-normal distribution. This could be related to
  the term with exponential of $-\operatorname{tr}[G]$ in Eq.~(\ref{eq:pdf-g}). Soize
  \cite{soize2000nonparametric} also gave the joint distribution of the random eigenvalues of
  $[\mathbf{R}]$ (see Eq.~(64) therein), which theoretically can be marginalized to yield the
  analytical distribution of $k$. The analytical expression is, however, extremely complicated and
  is not pursued in the present work.  


\begin{figure}[htbp]
  \centering
  \subfloat[PDF  of $\Delta C_3$, Case 1]{\includegraphics[width=0.42\textwidth]{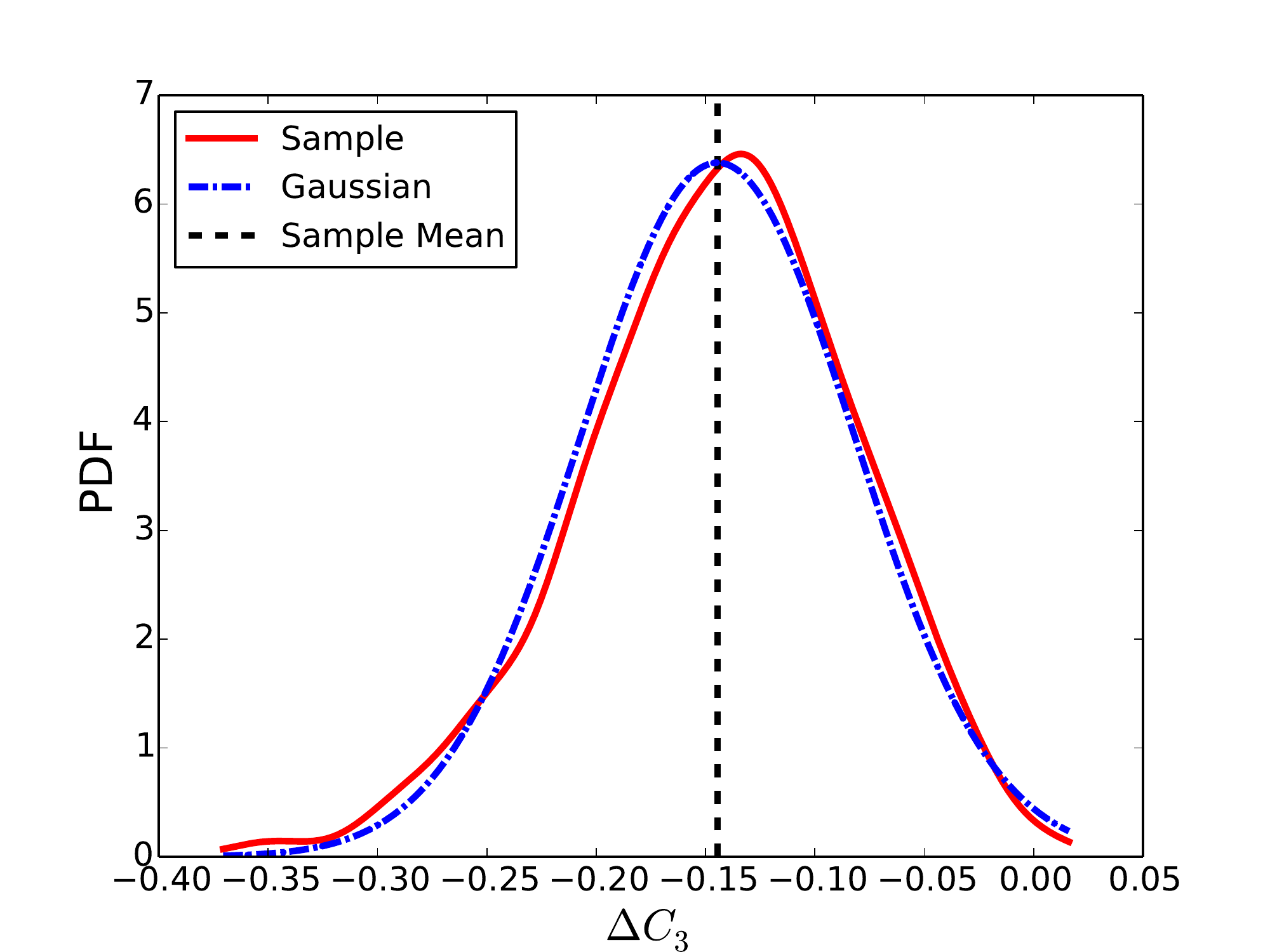}}
  \subfloat[PDF  of $\Delta C_3$, Case 2]{\includegraphics[width=0.42\textwidth]{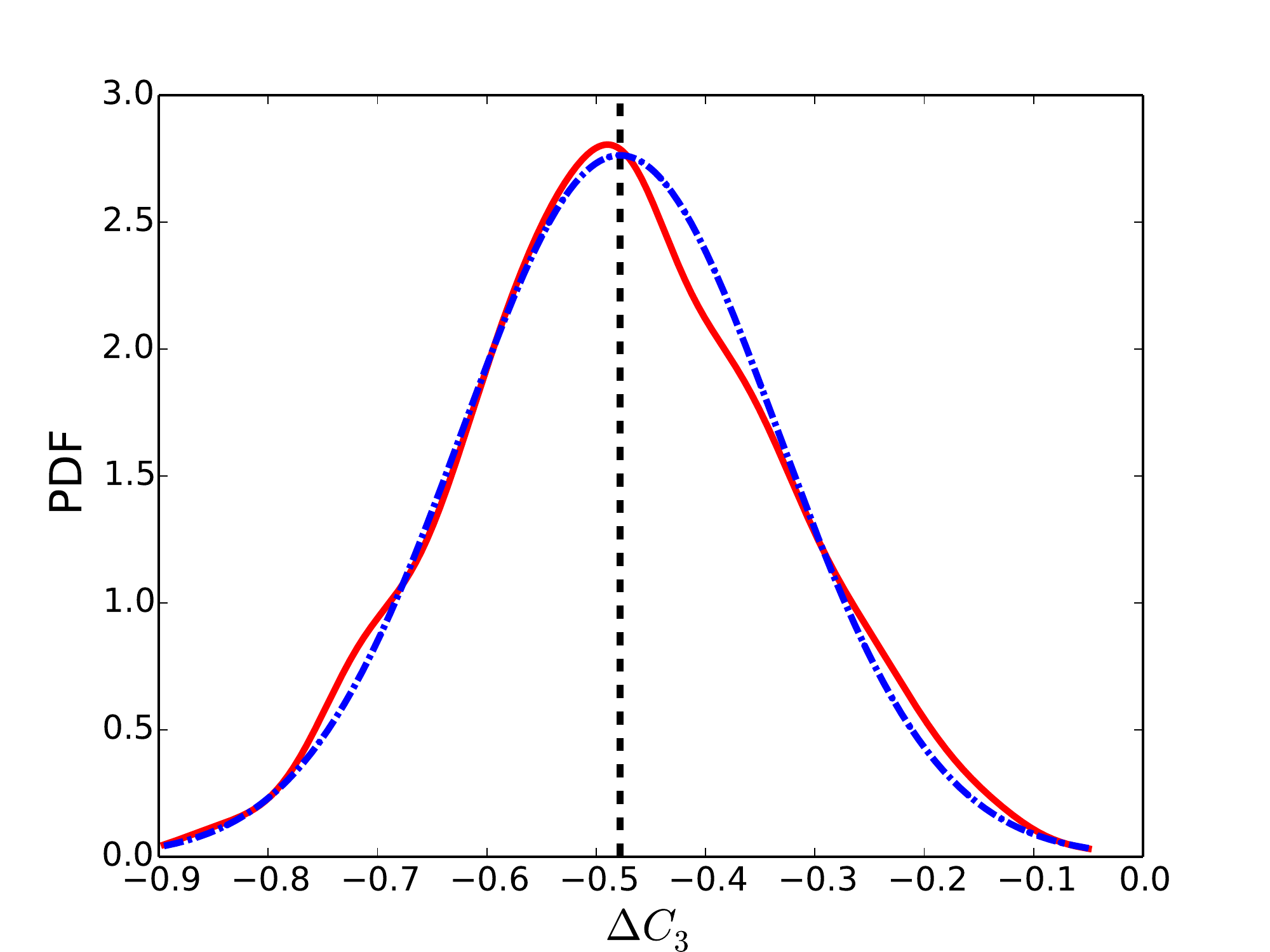}} \\
  \subfloat[PDF  of $\Delta \varphi_3$, Case 1]{\includegraphics[width=0.42\textwidth]{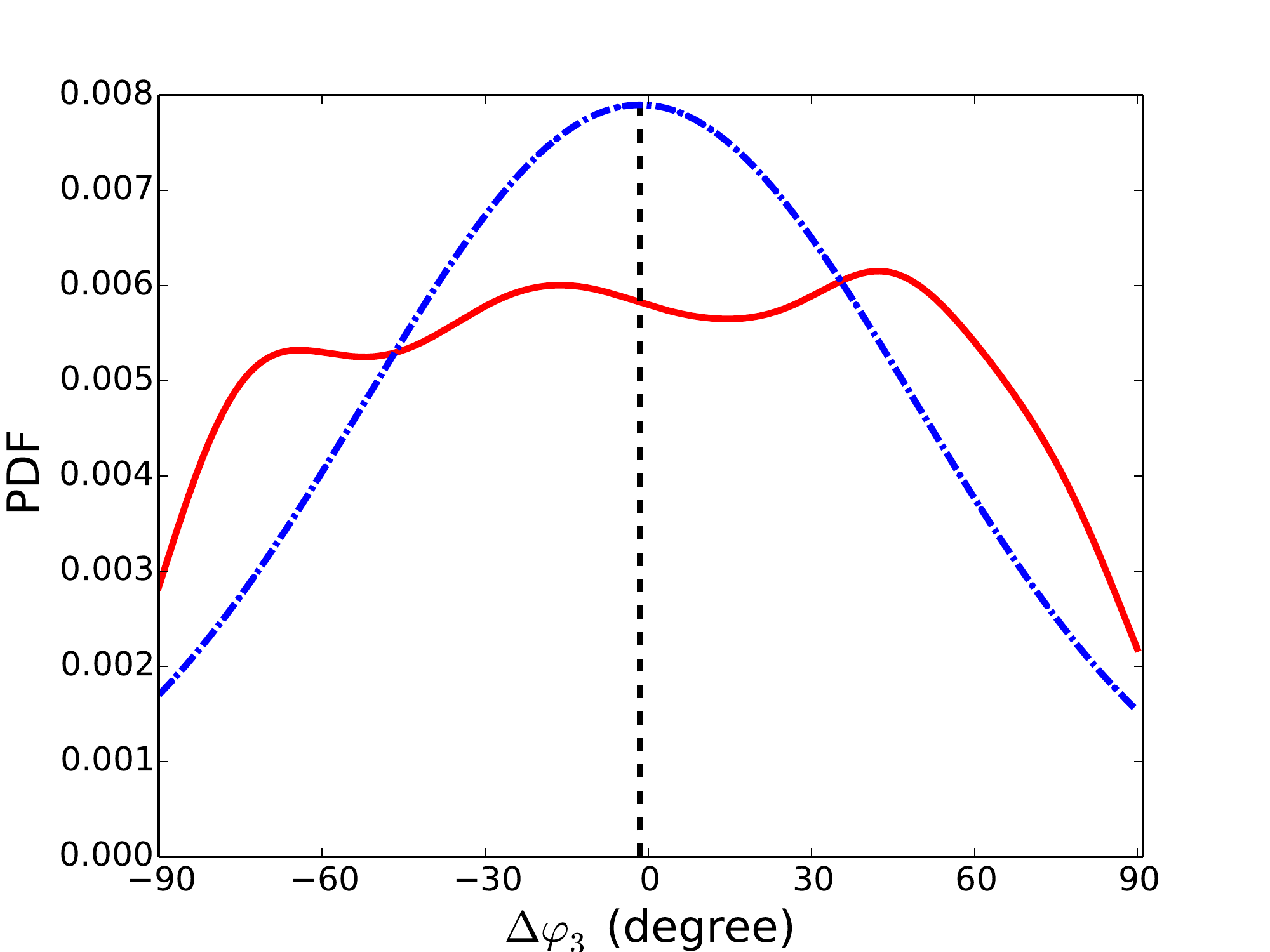}}
  \subfloat[PDF  of $\Delta \varphi_3$, Case 2]{\includegraphics[width=0.42\textwidth]{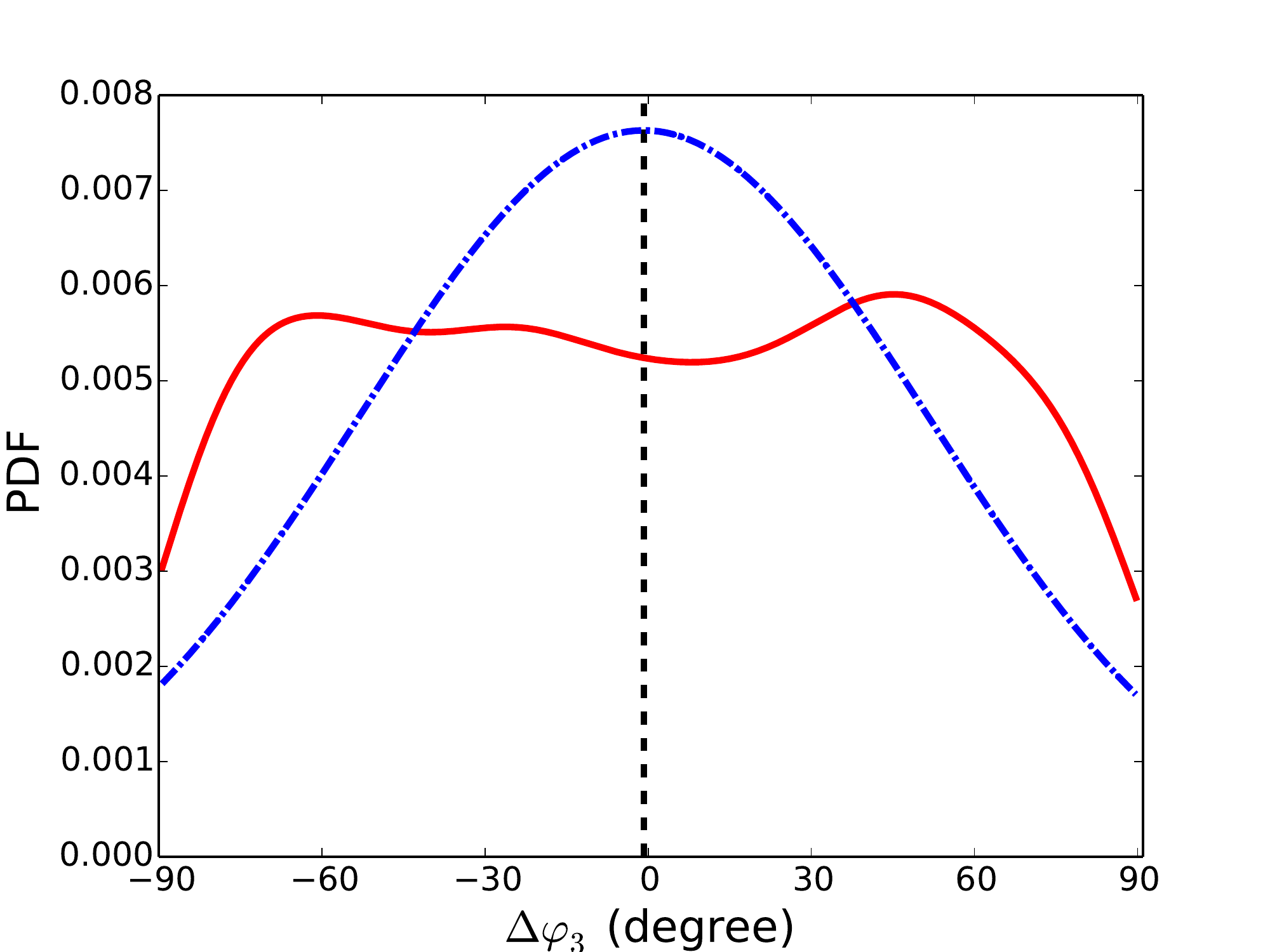}} \\
  \subfloat[PDF $\Delta \ln k$, Case 1]{\includegraphics[width=0.42\textwidth]{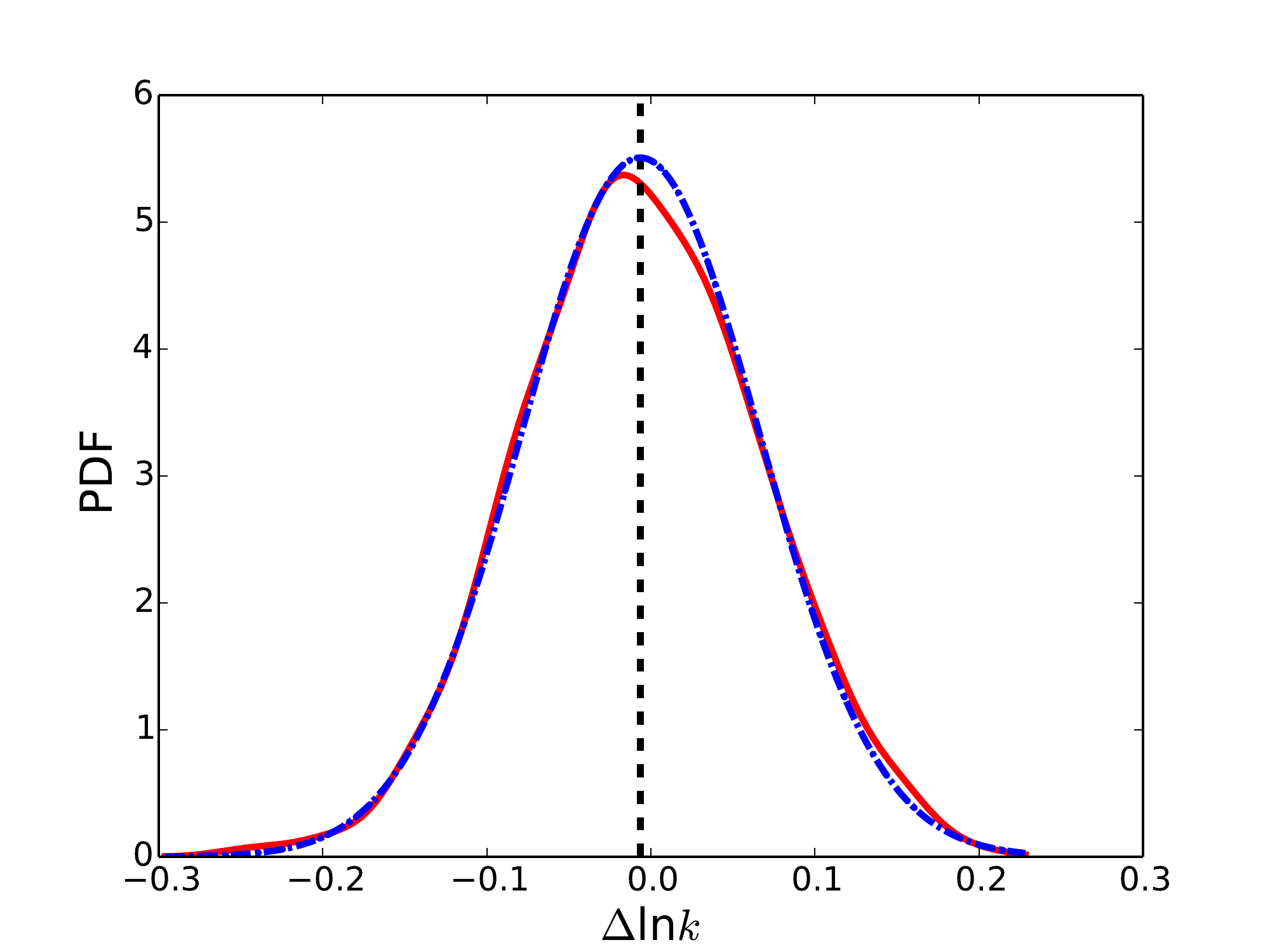}}
  \subfloat[PDF $\Delta \ln k$, Case 2]{\includegraphics[width=0.42\textwidth]{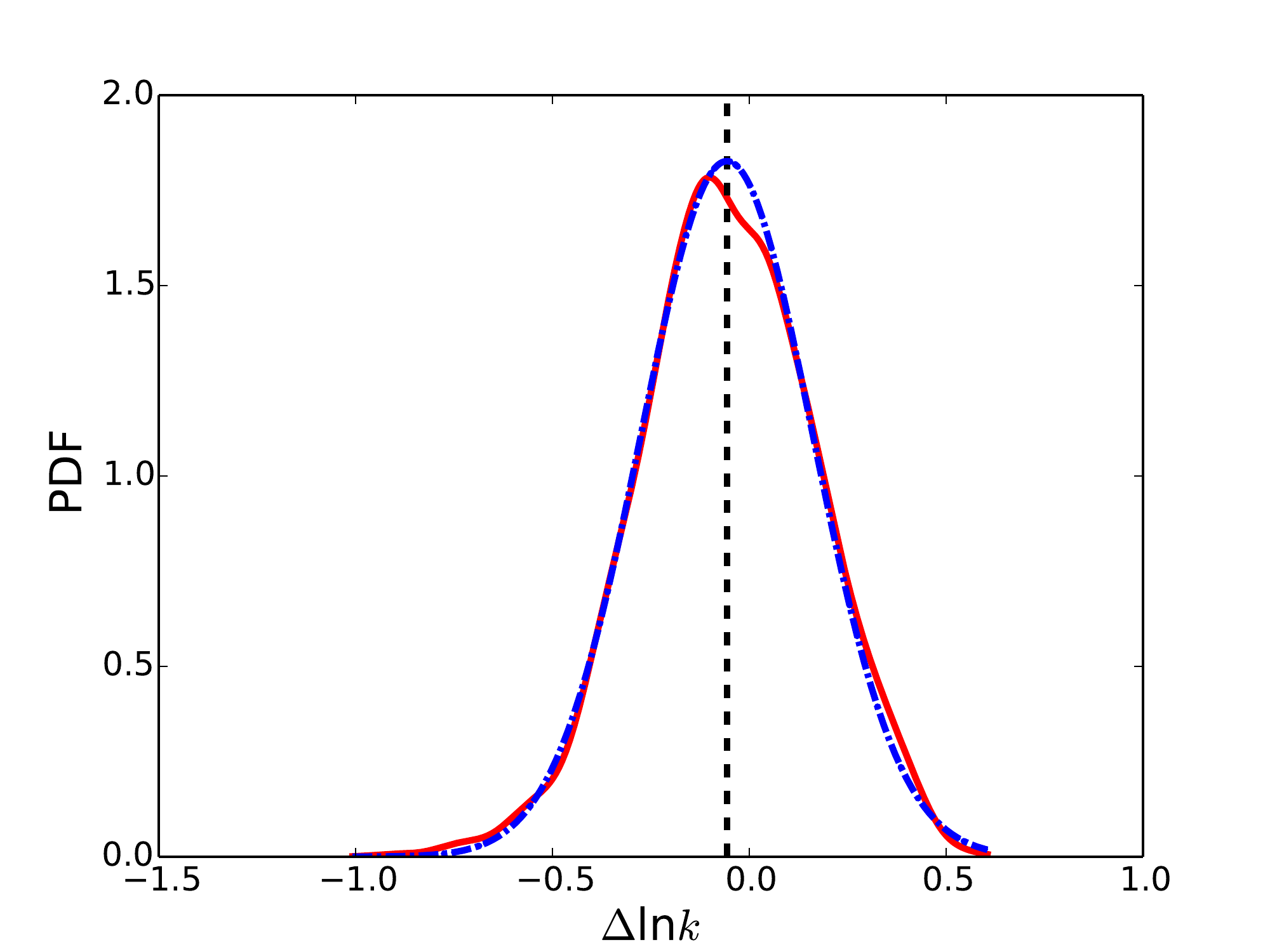}}      
   \caption{Probability density functions (PDF) of the perturbations ($\Delta C_3$, $\Delta
     \varphi_3$, and $\Delta \ln k$ ) in the physical variables for point B ($x/H=2.0, y/H=0.01$)
     located in the near-wall region.  The PDFs for Case 1 ($\delta = 0.2$) and Case 2 ($\delta =
     0.6$) are compared. }
   \label{fig:shapeLocB}
  \end{figure}

  It is also of interest to see the probability distribution of the Reynolds stress at a near-wall
  location with limiting-state turbulence. Therefore, similar to Figs.~\ref{fig:bayLocA}
  and~\ref{fig:shapeLocA}, the scatter plots and distribution of physical variable are presented in
  Figs.~\ref{fig:bayLocB} and~\ref{fig:shapeLocB}, respectively, for point B located very close to
  the bottom wall (at $x/H = 2.0$ and $y/H = 0.01$). The true Reynolds stress at this point is two
  dimensional, and thus it is located right on the bottom edge of the Barycentric triangle ($C_3 =
  0$) as indicated in Figs.~\ref{fig:bayLocB}a and~\ref{fig:bayLocB}b. Recall that $C_3$ indicates
  the degree of isotropy (three-dimensionality) of the turbulence. Hence, $C_3 =0$ indicates that
  the turbulence is far from isotropic (in fact, it is two-dimensional), which is because the
  blocking of the wall suppressed almost all fluctuations in wall-normal direction.  A standard eddy
  viscosity model, on the contrary, would predict a nearly isotropic turbulence state with $C_3 =
  1$, located near the top vertex of the Barycentric triangle (indicated as dark/red filled
  circles). The scatter plots corresponding to Cases 1 and 2 are shown in Figs.~\ref{fig:bayLocB}a
  and~\ref{fig:bayLocB}b, and the corresponding probability density contours are presented in
  Figs.~\ref{fig:bayLocB}c and~\ref{fig:bayLocB}d.  It can be seen that the sample distributions are
  influenced by the constraints in both cases, since the baseline state is located right next to the
  boundaries. Moreover, the difference between the sample mean and the baseline are large in both
  cases, which is also attributed to the constraints as discussed above.  The distributions of
  $\Delta C_3$, $\Delta \varphi_3$, and $\Delta \ln k $ are presented in
  Figure~\ref{fig:shapeLocB}. For both Cases 1 and 2, the sample distributions of $\Delta C_3$ and
  $\Delta \varphi_3$ deviate significantly from Gaussian, while the distribution of $k$ is still
  quite close to log-normal.

\begin{figure}[htbp]
  \centering
  \hspace{2em}\includegraphics[width=0.7\textwidth]{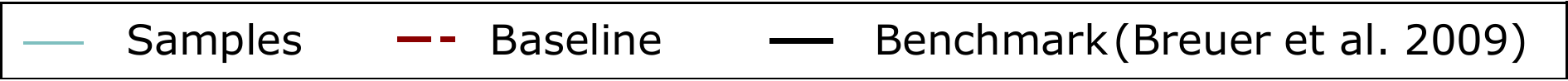}
   \subfloat[Samples of $R_{12}$ profiles, Case 1]{\includegraphics[width=0.8\textwidth]{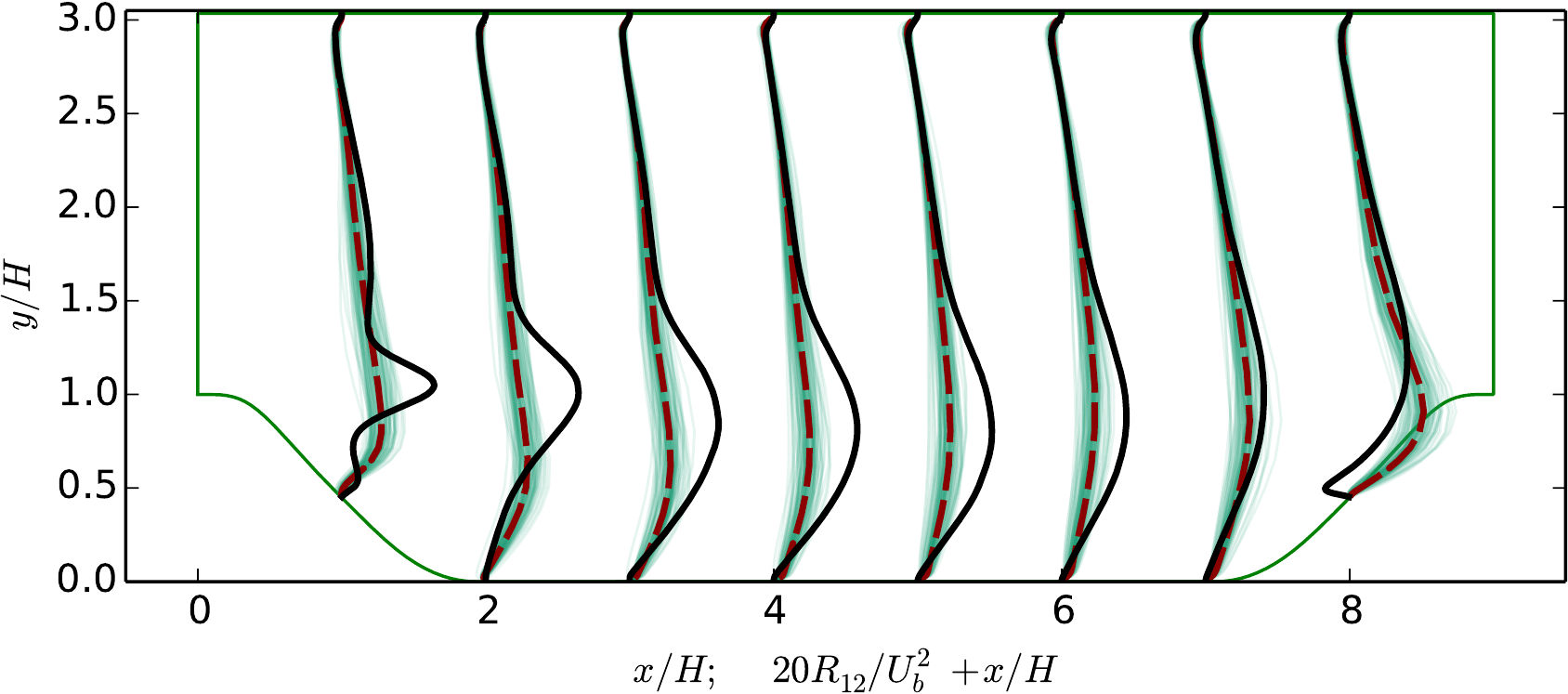}}\\
   \subfloat[Samples of $R_{12}$ profiles, Case 2]{\includegraphics[width=0.8\textwidth]{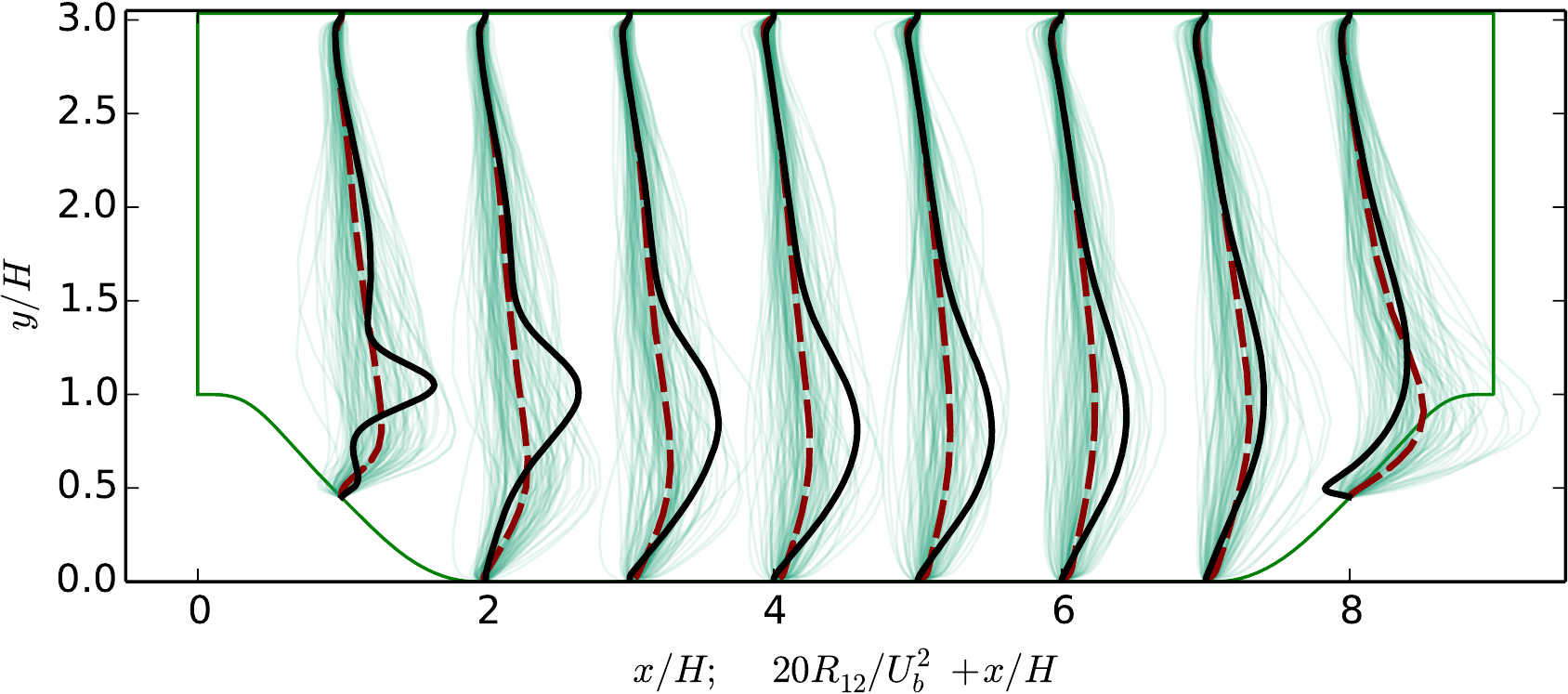}}      
   \caption{Comparison sample profiles $R_{12}$ for Case 1 and Case 2. The profiles are shown at
     eight streamwise locations $x/H = 1, \ \cdots, 8$, compared
     to the baseline results and the benchmark results obtained by direct numerical simulation~\cite{breuer2009flow}.}
  \label{fig:Tauxy}
\end{figure}

\begin{figure}[htbp]
  \centering
  \hspace{2em}\includegraphics[width=0.7\textwidth]{./pehill-U-legend-dns-noObs}
  \subfloat[Samples of $k$ profiles, Case 1]{\includegraphics[width=0.8\textwidth]{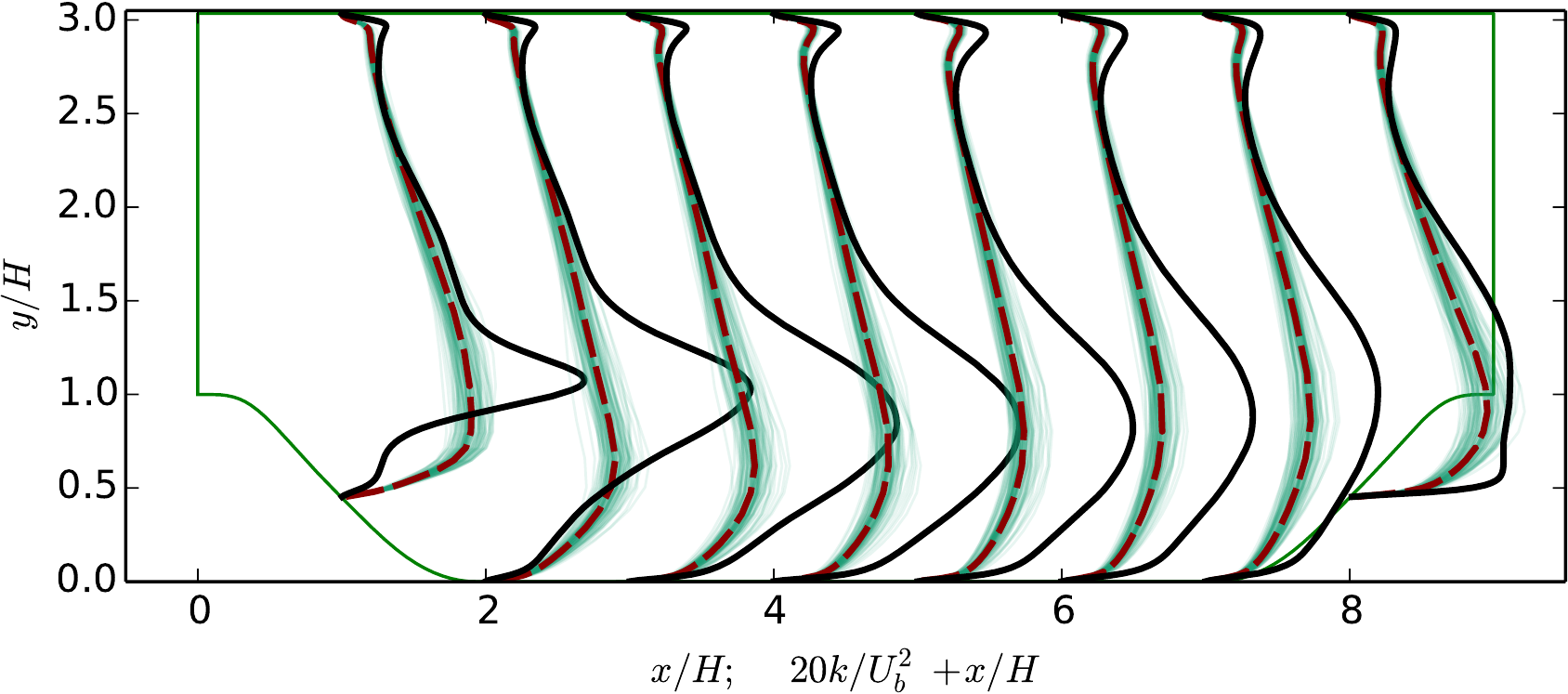}}\\
  \subfloat[Samples of $k$ profiles, Case 2]{\includegraphics[width=0.8\textwidth]{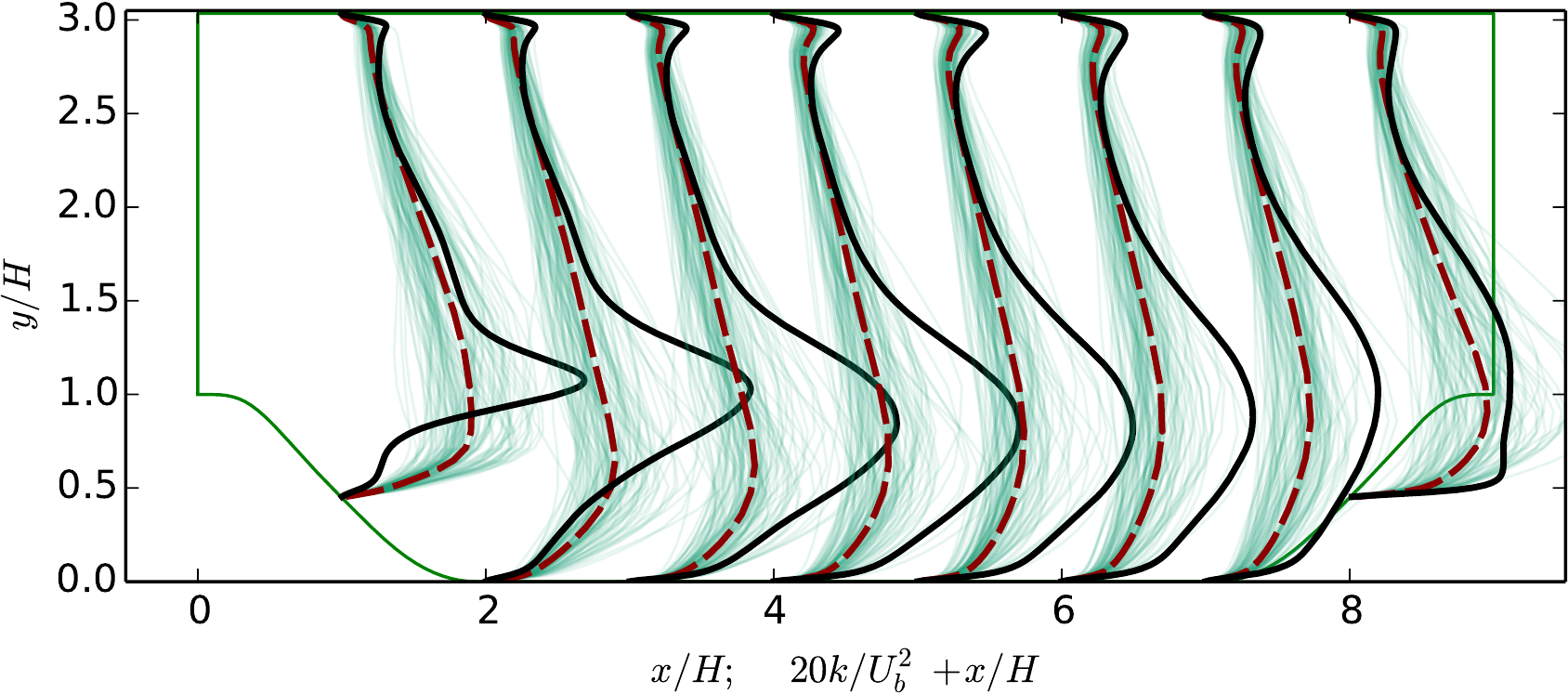}}      
   \caption{Comparison of sample profiles of turbulence kinetic energy $k$ profiles for Case 1 and
     Case 2. The profiles are shown at eight streamwise locations $x/H = 1, \ \cdots, 8$, compared
     to the baseline results and the benchmark results obtained by direct numerical simulation~\cite{breuer2009flow}.}
  \label{fig:TKEc}
\end{figure}

  The analysis above focused on the marginal distributions of the Reynolds stresses at two
  representative locations with a generic and a limiting state turbulence state. Since the Reynolds
  stress is modeled as a random matrix field, we present the turbulent shear stress $R_{12}$ and the
  turbulent kinetic energy $k$ in Figs.~\ref{fig:Tauxy} and~\ref{fig:TKEc}, respectively. These two
  quantities are the most relevant for the flow over period hills investigated here. The samples at
  eight streamwise locations $x/H = 1, 2, \cdots, 8$ along with the benchmark and the baseline
  results. The geometry of the physical domain is also plotted to facilitate visualization.  The
  sample mean profiles coincide with the baselines in most regions for all cases, and thus they are
  omitted to avoid cluttering. This is in contrast to the deviation of the sample means from the
  baseline results in the Barycentric triangle as observed in Figs.~\ref{fig:bayLocA} and
  \ref{fig:bayLocB} for the same cases.

It can be seen that the turbulent shear stress $R_{12}$ in the baseline and benchmark results agree
quite well with each other in the upper channel ($y/H = 2$ to 3) but deviate dramatically in the
free-shear and recirculation regions.  In the case with a small dispersion parameter $\delta = 0.2$,
Fig.~\ref{fig:Tauxy}a shows that the uncertainty range as represented by the $R_{12}$ samples is not
able to cover the benchmark results in regions where large discrepancies exist between the baseline
and benchmark. This indicates that the prior may be overly confident, which may lead to difficulties
in Bayesian inferences. With a large dispersion parameter $\delta = 0.6$ in Case 2, the scattering
range of $R_{12}$ increases and covers the benchmark in the entire domain, which is shown in
Fig.~\ref{fig:Tauxy}b.  However, since the dispersion parameter is spatially uniform, it also has
large scattering in the upper channel, where in fact the discrepancies between baseline and
benchmark results are rather small.  Similar patterns are also observed in the turbulent kinetic
energy $k$ profiles, which are shown in Figs.~\ref{fig:TKEc}a and~\ref{fig:TKEc}b for Cases 1 and 2,
respectively. However, a notable difference is that the range of samples is not able to cover the
benchmark results even in Case 2 with a large discrepancy parameter. This is because the turbulent
kinetic energy $k$ is significantly underestimated in the baseline turbulence model, which is a
known deficiency of turbulence models applied this flow. The deficiency stems from the fact that the
shear stresses and consequently the modeled turbulent kinetic energy production are weaker than the
physical counterparts~\cite{billard2012development}.

\begin{figure}[!htbp]
  \centering
  \hspace{2em}\includegraphics[width=0.7\textwidth]{./pehill-U-legend-dns-noObs} 
  \subfloat[Sample of $R_{12}$ profiles, Case 3]{\includegraphics[width=0.8\textwidth]{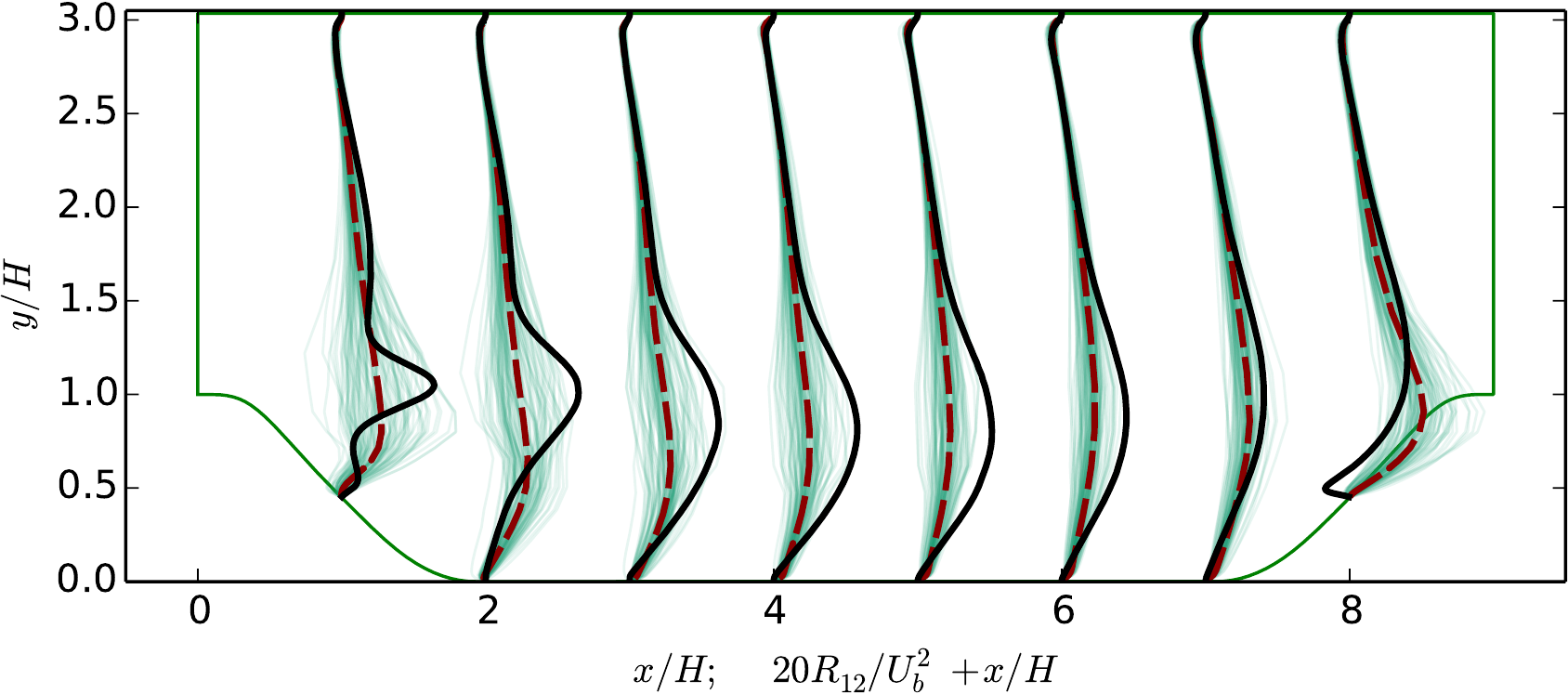}}\\
  \subfloat[Sample of $k$ profiles, Case 3]{\includegraphics[width=0.8\textwidth]{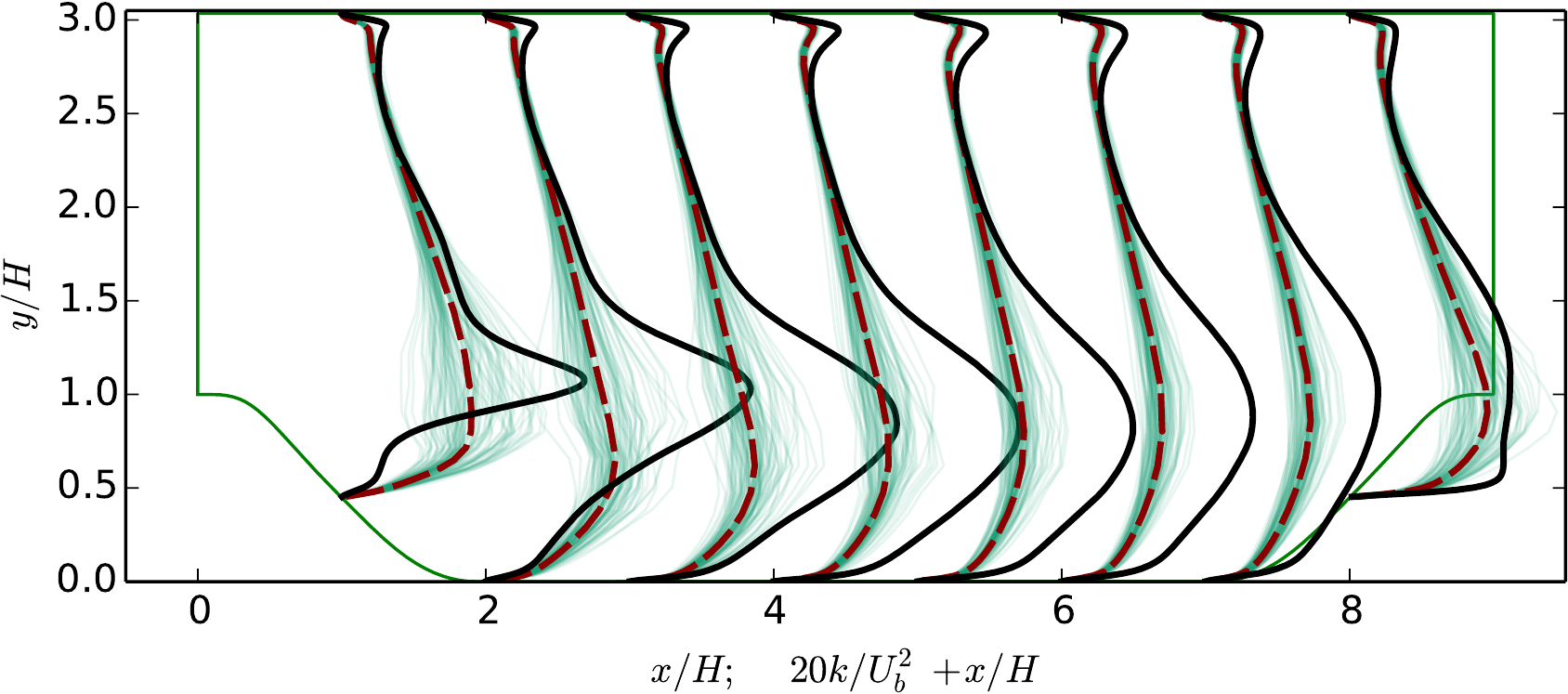}}
  \caption{The sample profiles of turbulent shear stress $R_{12}$ and turbulent kinetic energy $k$
    for Case 3. The ensemble profiles are shown at eight streamwise locations $x/H = 1, \cdots, 8$,
    compared to the baseline results and the benchmark results obtained by direct numerical simulation~\cite{breuer2009flow}.}
  \label{fig:Rnon}
\end{figure}

In light of the comparison between the samples and the benchmark above, it can be seen that using a
spatial uniform dispersion parameter $\delta$ leads to scattering larger than necessary in the upper
channel, but with insufficient scattering in regions that are indeed problematic for the baseline
model (e.g., the free shear and recirculation regions).  It is known from experiences in turbulence
modeling that standard turbulence models have difficulties in these regions, using a nonuniform,
informative dispersion field $\delta(x)$ as shown in Fig.~\ref{fig:domain_pehill} is justified in
Section~\ref{sec:setup} above. The samples of $R_{12}$ and $k$ are presented in
Fig.~\ref{fig:Rnon}. In both cases the scattering of the samples are larger in the free-shear and
recirculation regions and are smaller in the upper channel, which is more consistent with the actual
discrepancies between the baseline and the benchmark results. A minor exception is that the
turbulent kinetic energy near the upper wall is underpredicted by the baseline model but this is not
reflected in the sample scatter, since this feature is not reflected in the prior knowledge in the
dispersion parameter $\delta(x)$ field.

Overall speaking, all sample profiles of the Reynolds stress component $R_{12}$ and the turbulent
kinetic energy $k$ shown in Figs.~\ref{fig:Tauxy}--\ref{fig:Rnon} are physically reasonable. These
samples are qualitatively similar to those obtained with the physics-based approach by directly
perturbing the physical variables $C_1, C_2$, and $\ln k$ around the baseline
results~\cite{xiao-mfu}.

\begin{figure}[!htbp]
  \centering
   \hspace{2em}\includegraphics[width=0.7\textwidth]{./pehill-U-legend-dns-noObs}
   \subfloat[Velocity samples, Case 1]{\includegraphics[width=0.8\textwidth]{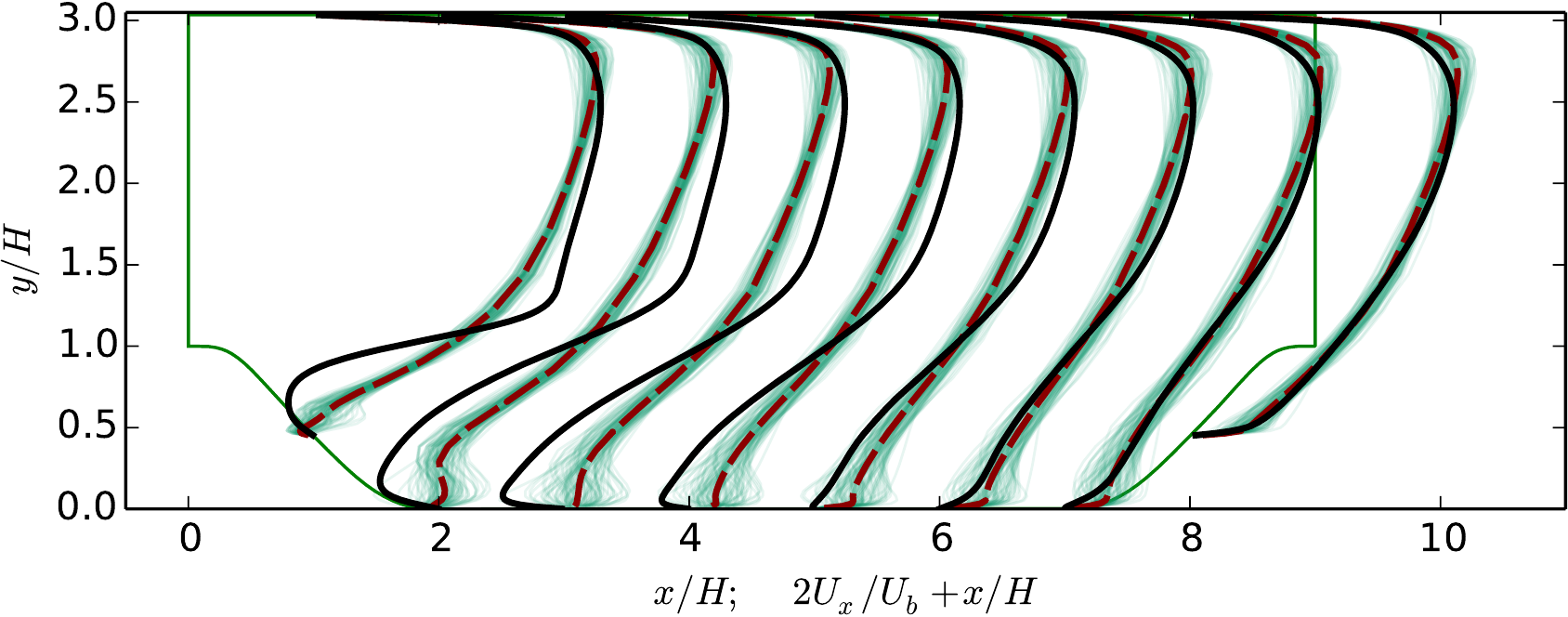}}\\
   \subfloat[Velocity samples, Case 2]{\includegraphics[width=0.8\textwidth]{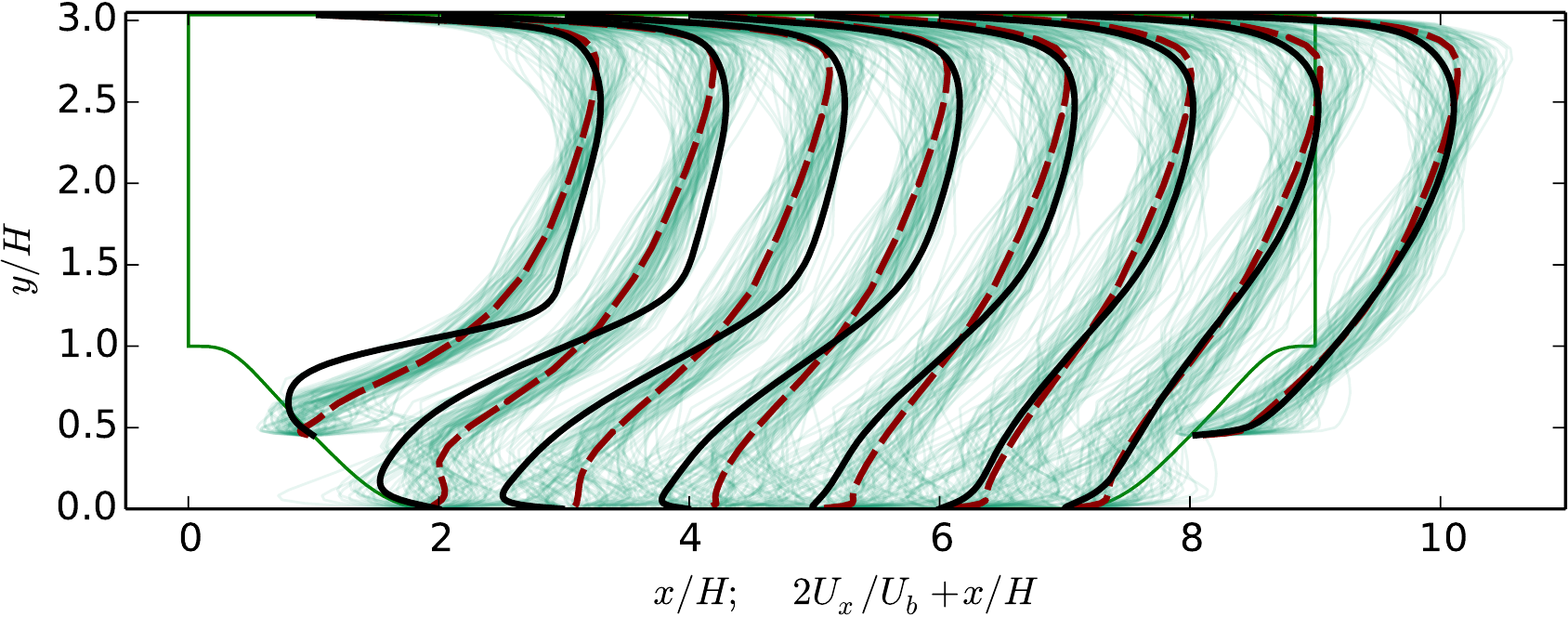}} \\
   \subfloat[Velocity samples, Case 3]{\includegraphics[width=0.8\textwidth]{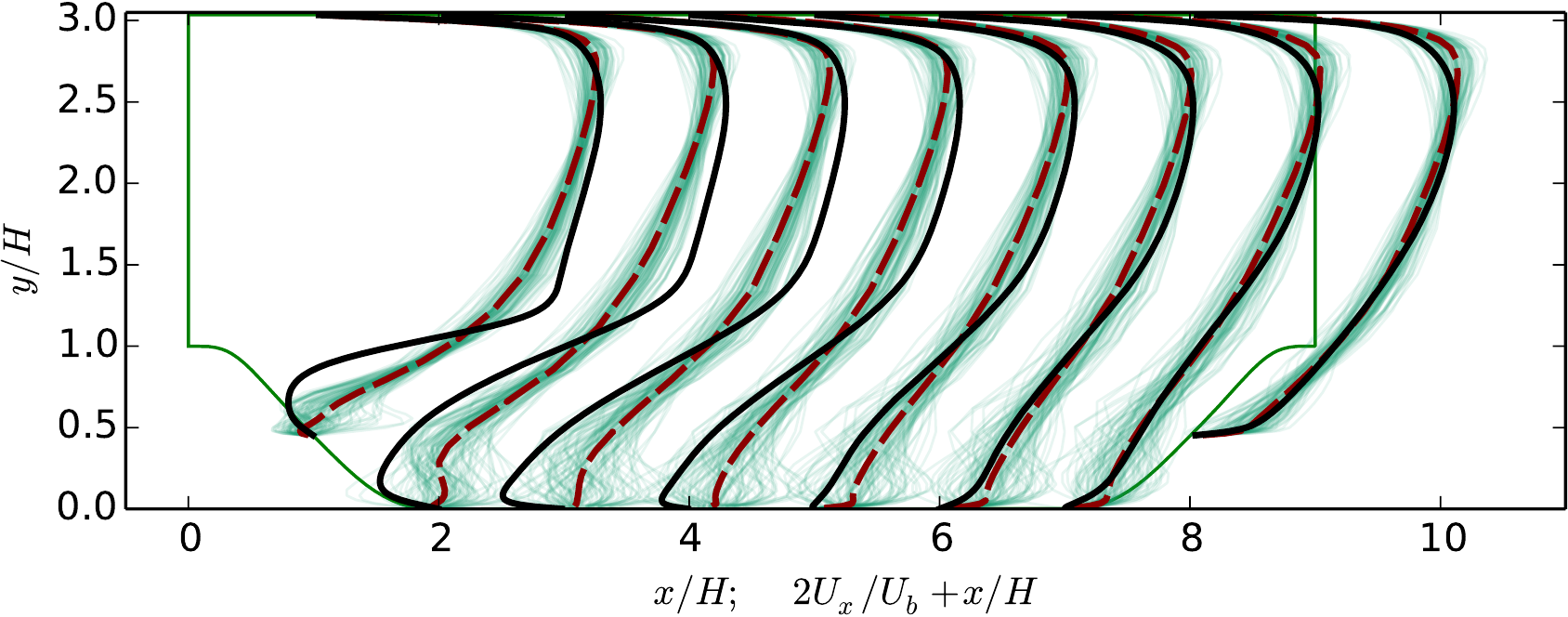}}\\
   \caption{Comparison of propagated velocity profile samples for Cases 1, 2, and 3. The profiles
     are shown at eight streamwise locations $x/H = 1, \cdots, 8$, 
     compared to the baseline results and the benchmark results obtained by direct numerical simulation~\cite{breuer2009flow}.}
  \label{fig:velocity}
\end{figure}

Ultimately it is the velocity fields and the associated auxiliary quantities (e.g., wall shear
stress, reattachment point, and pressure drop across the channel) that are of interest in turbulent
flow simulations. Therefore, the sampled Reynolds stress are propagated to velocities by using the
RANS solver tauFoam~\cite{xiao-mfu}, and the reattachment points are obtained by post-processing the
velocity fields.  The velocity profiles obtained for the three cases with different dispersion
parameters are presented and compared in Fig.~\ref{fig:velocity}.  Similar to the profiles of
$R_{12}$ and $k$, the velocity sample also show a small scattering in Case 1 ($\delta = 0.2$) and a
larger scattering in Case 2 ($\delta = 0.6$). The advantage of using an informative $\delta(x)$
field can also be observed in this figure. Compared to the uniformly small scattering in Case 1
(panel a) and uniformly large scattering in Case 2 (panel c), the velocity scattering in Case 3
(panel c) is small in the upper channel and large in the free shear and recirculation regions,
adequate to cover the benchmark results in both regions. Finally, we note that in all three cases
the velocity scattering in the lower part, particularly in the recirculation region, of the channel
is larger, even though the Reynolds stress scattering is spatially uniform in Cases 1 and 2. This
observation suggests that the velocity in the recirculation region is more sensitive to the Reynolds
stresses, an important physical insight that is provided by the RANS solver. Similar analyses have
been performed on other quantities of interest including the bottom wall shear stress and the
reattachment point. Qualitatively similar observations are made as those presented above for the
velocities, and thus the figures for the other QoIs are omitted here.


\section{Discussions}
\label{sec:discuss}

 Whether the obtained maximum entropy distribution of Reynolds stress~$[\mathbf{R}]$ 
	should have zero measure on the two-component (and one-component) limiting
  states is an issue open for discussion. We explain below the implications based on the choices
  made in the proposed framework. Although practically we did not encounter this difficulty in
  this study, the discussion below adds to the theoretical completeness.  The probability
  distribution for the normalized random matrix $[\mathbf{G}]$ as given in Eq.~(\ref{eq:pdf-g}) has
  zero measure on the subset $\mathbb{M}_d^{+0}\setminus\mathbb{M}_d^{+}$ of singular positive
  semidefinite matrices, since $\det([G]) = 0$ in this set and thus $ p_{[\text{\textbf{\tiny
        G}}]}([G])|_{[G] \in \mathbb{M}_d^{+0} \setminus \mathbb{M}_d^{+}} = 0$. This is analogous
  to the $\chi^2$-distribution in the scalar case with more than three degrees of freedom.  However,
  a valid distribution for semi-definite matrices could have a concentration of measure in the set
  of singular matrices, i.e., it is possible that $ p_{[\text{\textbf{\tiny R}}]}([R])|_{[R] \in
    \mathbb{M}_d^{+0} \setminus \mathbb{M}_d^{+}} \neq 0 $. One can draw an analogy to the
  exponential distribution, for instance, in the scalar case. If the specified mean
  $[\underline{R}]$ of $[\mathbf{R}]$ is singular, i.e., $\det([\underline{R}])$ = 0, then
  $\det([\underline{L}_R]) = 0$. Consequently, any realization $[R]$ constructed from
  Eq.~(\ref{eq:a-transform}), $[\mathbf{R}] = [\underline{L}_R]^T [\mathbf{G}] [\underline{L}_R]$,
  is also singular since the distributive nature of determinants leads to $\det([R]) = 0$.  The set
  $\mathbb{M}_d^{+0}\setminus\mathbb{M}_d^{+}$ of singular Reynolds stresses maps to the
  two-component limit edge ($C_3 = 0$) of the Barycentric triangle. This mapping can be explained by
  the fact that all matrices in this set have at least one eigenvalue that is zero (as they have zero
  determinants), which corresponds to two-component turbulence.  This correspondence is indicated in
  Fig.~\ref{fig:bary}a.  The physical interpretation of the mathematical consequences above can be
  summarized as follows:
  \begin{enumerate}
  \item When the specified Reynolds stress mean $[\underline{R}]$ corresponds to a turbulence state
    in a two-component limit, the samples of perturbed Reynolds stresses also have two-component
    states;
\item When the specified mean corresponds to a generic three-component state, the samples can
  becomes infinitely close to a two-component limit but has zero probability of falling on the
  limit.
  \end{enumerate}

\section{Conclusion}
In this work we propose a random matrix approach to guarantee Reynolds stress realizability for
quantifying model-form uncertainty in RANS simulation.  In this approach, the Reynolds stress field
is described with a probabilistic model of a random field of positive semidefinite matrices with
specified mean and correlation structure.  The marginal probability distribution of the Reynolds
stress at any particular location satisfies the maximum entropy principle. To sample such a random
matrix field, Gaussian random fields with specified covariance kernel are first generated and then
are mapped to the field of positive semidefinite matrices based on polynomial chaos expansion and
reconstruction. Numerical simulations have been performed with the proposed approach by sampling the
Reynolds stress and propagating through the RANS solver to obtain velocities. The simulation results
showed that generated Reynolds stress fields not only have the specified statistics, but also are
physically reasonable.  The mathematically rigorous approach based on random matrix and maximum
entropy principle is a promising alternative to the previously proposed physics-based approach for
quantifying model-form uncertainty in RANS simulations. Moreover, it can be used to gauge the
entropy of the prior distributions specified in the physics-based approach and thus to measure the
information introduced therein. Detailed comparison between the two approaches will be pursued as
follow-on work in the future.






\appendix
\section{Summary of the Algorithm of the Proposed Method}
\label{sec:summary}
Given the mean Reynolds stress field $[\underline{R}](x)$ along with the correlation function
structure of the random matrix field $[\mathbf{L}](x)$ as specified in Eq.~(\ref{eq:Lx}), the
following procedure is performed:

\begin{enumerate}
\item  Expansion of given marginal distributions and covariances kernels:
  \begin{enumerate}[label*=\arabic*.]
  \item Perform the Cholesky factorization of the mean Reynolds stresses $[\underline{R}]$ at each
    cell according to Eq.~(\ref{eq:lu-decomp}), which yields field $\underline{L}_R(x)$ of upper
    triangular matrices.
  \item Perform KL expansion for the kernel function by solving the Fredholm equation~(\ref{eq:kl}) to
    obtain eigenmodes $\sqrt{\lambda_\alpha} \; \phi_\alpha(x)$.
  \item For off-diagonal terms only, perform polynomial expansion of the marginal PDF as
    described in Eq.~(\ref{eq:v-pdf}) at each cell. Coefficients $U_\beta$ are obtained from
    Eq.~(\ref{eq:U_i}), where $\beta = 1, \cdots, N_p$, and $N_p$ is the number of polynomials
    retained in the expansion.
  \end{enumerate}

\item Sampling and reconstruction of random matrix fields for Reynolds stresses:

\begin{enumerate}[label*=\arabic*.]
\item For each element $\mathbf{L}_{i j}$ of the random matrix field $[\mathbf{L}]$, independently
  draw $N_{\textrm{KL}}$ sample from the standard Gaussian distribution $\omega_{i j, \alpha}$ where
  $\alpha = 1, \cdots, N_{\textrm{KL}}$, e.g., with random sampling or Latin hypercube sampling
  method~\cite{Helton:2003fc}.

\item Synthesize realizations of the off-diagonal terms based on KL expansion:
  \begin{align}
    \mathbf{w}_{ij}(x) & = \sum_{\alpha=1}^{N_{\textrm{KL}}} \, \sqrt{\lambda_\alpha} \,
    \phi_\alpha(x) \;    \boldsymbol{\omega}_\alpha   \quad \textrm{with} \quad  i < j \notag \\ 
    \mathbf{L}_{i j}(x) & = \sigma_d \mathbf{w}_{ij}(x) \notag
  \end{align}

\item Synthesize the realizations of the diagonal terms based on KL and PCE expansions:
  \begin{equation}
    \mathbf{u}_{i}(x) = \sum_{\beta = 0}^{N_p} U_\beta(x) \Psi_\beta(\mathbf{w}_{ii}(x))   \notag
  \end{equation}
  where the Gaussian random field sample $\mathbf{w}_{ii}(x)$ obtained in the previous step is used.
\item Synthesize the diagonal terms of matrix $[\mathbf{L}]$ from $\mathbf{L}_{i i}(x) = \sigma_d
  \sqrt{\mathbf{u}_{i}}$, where $i = 1, 2, 3$.
\item Reconstruct $[\mathbf{G}]$ from $[\mathbf{G}] = [\mathbf{L}]^T [\mathbf{L}]$ and then
  reconstruct the random matrix $[\mathbf{R}]$ from $[\mathbf{R}] = [\underline{L}_R]^T [\mathbf{G}]
  [\underline{L}_R]$.
\end{enumerate}

\item Propagation the Reynolds stress field through the RANS solver to obtain velocities and
    other QoIs:

\begin{enumerate}[label*=\arabic*]
\item Use the obtained sampled Reynolds stress to velocity and other QoIs by solving the RANS
  equation.
\item Post-process the obtained velocity and QoI samples to obtain statistical moments.
\end{enumerate}

\end{enumerate}

\section{Nomenclature}
\label{app:notation}
\begin{tabbing}
  00000000000\= this is definition\kill 
  {\emph{Subscripts/Superscripts}}{}\\
  $i$, $j$ \> tensor indices ($i, j = 1, 2, 3$).  Repeated indices do not
  imply summation.  \\
  $\alpha$, $\beta$ \> indices for terms in Karhunen--Loeve  and  polynomial chaos expansions \\ 

  \\ 
  \emph{Sets, operators, and decorative symbols} \>  {} \\
  $\operatorname{Cov} (\cdot, \cdot)$  \> covariance of two random variables \\
  $\operatorname{det}(\cdot)$ \> determinant of a matrix \\
  $\mathbb{E}\{ \cdot \}$ \> expectation of a random variable \\
  $\mathcal{GP}$ \> Gaussian spatial random process (random field) \\

  $\mathbb{M}_d^{s}$ \> the set of all $d \times d$ symmetric matrices \\
  $\mathbb{M}_d^{+}$ \> the set of all $d \times d$ symmetric, positive definite matrices \\
  $\mathbb{M}_d^{+0}$ \> the set of all $d\times d$ symmetric, positive semi-definite matrices \\
  $\operatorname{tr}(\cdot)$ \> trace of a matrix \\
  $\operatorname{Var} (\cdot, \cdot)$  \> variance of a random variable \\
  $\ln$ \> the natural logarithm \\
  $\mathbbm{1}_{\Box}(\cdot)$  \> indicator function; takes value one if $\cdot \in \Box$  \\
  $\vec{\cdot}$ \>  vector \\
  $[\cdot]$ \>  matrix \\
  $\underline{\Box}$ \> mean value of variable $\Box$ \\
  $\langle \cdot \rangle$ \> ensemble average/expectation of a random variable \\
  $\| \cdot \|_F$  \> Frobenius norm \\
  $\sum$ \> summation \\
  $\prod$ \> product \\
  \\
  \emph{Roman letters} \>  {} \\
  $[A]$  \> anisotropy tensor of the Reynolds stress \\
  $C_{[G]}$  \> normalization constant in the PDF of random matrix $[G]$ \\
  $C_{1}, C_{2}, C_{3}$  \> Barycentric coordinates \\
  $d$ \> dimension of matrices ($d=3$ implied unless noted otherwise) \\
  $\vec{e}_1, \vec{e}_2, \vec{e}_3$ \> eigenvectors of tensor $[A]$ \\
  $[E]$ \>   matrix formed by the eigenvectors of $[A]$, i.e.,  $[E] = [\vec{e}_1, \vec{e}_2, \vec{e}_3]$ \\
  $[\mathbf{G}], [G]$ \> a positive definite matrix with identity matrix $[I]$ as its mean \\ 
  $[I]$ \>  identity matrix \\
  II, III \>  second and third invariants of tensor ($3 \times 3$ matrix) $[A]$ \\
  $K$ \> kernel of a Gaussian process \\
  $[\mathbf{L}], [L]$ \> an upper triangle matrix (e.g., obtained with Cholesky factorization) \\
  $l$ \> length scale of Gaussian process \\
  $\mathbf{N}$  \> Gaussian normal distribution \\
  $N_p$ \> order of polynomial in the polynomial chaos expansion \\
  $N_{\textrm{KL}}$ \> number of modes in KL expansion \\
  $p$ \> probability density function \\
  $[R]$ \> (negative of) Reynolds stress tensor with $R_{ij} = \langle u_i' u_j' \rangle$  \\
  $S(\cdot)$ \> entropy measure of a PDF \\
  $\mathbf{u}$ \> random variable used to define the diagonal terms of matrix $[\mathbf{L}]$ \\
  $\mathbf{v}_i$ \> velocity at a particular location (random variable) \\
  $\underline{v}_i$ \> mean velocity \\
  $\mathbf{v}_i'$ \>  the $i$\textsuperscript{th} component of the fluctuation velocity (random
  variable) \\
  $\mathbf{w}$ \> standard Gaussian random variable, i.e., $\mathbf{w} \sim \mathbf{N}(0, 1)$ \\
  $z, t$ \> auxiliary variables \\
  \\
  {\emph{Greek letters}}{}\\
  $\delta$ \> dispersion parameter (uncertainty of the random matrix) \\
  $\nu$ \> variable indicating the order of the matrix \\
  $\Gamma$ \> Gamma function \\
  $\tilde{\lambda}_i$ \>  eigenvalues  of the turbulence anisotropy tensor $[A]$ \\
  ${\Lambda}$ \>   $3 \times 3$ diagonal matrix  with eigenvalues of $[A]$ as elements, i.e.,
  $\Lambda = [\tilde{\lambda}_1, \tilde{\lambda}_2, \tilde{\lambda}_3]$ \\
  $\lambda_\alpha, \phi_\alpha$ \> eigenvalues and basis functions of $\alpha$\textsuperscript{th} mode obtained from KL expansion \\
  $\Psi_\beta$ \> $\beta$\textsuperscript{th} order Hermite polynomial of the standard Gaussian random
  variable $\mathbf{w}$ \\
  $\Omega$ \> spatial domain of the fluid flow field
 \end{tabbing}


\end{document}